\documentclass[a4paper,11pt]{article}
\pdfoutput=1 

\usepackage{jinstpub} 

\newcommand{\fins}{f}
\newcommand{\CW}[1]{\textcolor{black}{#1}} 

\begin{document}


\title{Cryogenic bath-type heat exchangers for ultra-pure noble gas applications}


\author[a,b,1]{M. Murra,\note{Corresponding author}}
\author[a,1]{D. Schulte,}
\author[c]{I. Cristescu,}
\author[d]{J.-M. Disdier,}
\author[a]{C. Huhmann,}
\author[e]{D. Tatananni,}
\author[a]{C. Weinheimer}

\affiliation[a]{Institute for Nuclear Physics, University of M\"unster, \\Münster, Germany}
\affiliation[b]{Physics Department, Columbia University,\\New York, USA}
\affiliation[c]{Tritium Laboratory Karlsruhe (TLK), Karlsruhe Institute of Technology (KIT),\\ Eggenstein-Leopoldshafen, Germany}
\affiliation[d]{SUBATECH, IMT Atlantique, CNRS/IN2P3, Universit´e de Nantes,\\Nantes, France}
\affiliation[e]{INFN-Laboratori Nazionali del Gran Sasso,\\L’Aquila, Italy}


\emailAdd{michaelmurra@uni-muenster.de}
\emailAdd{DennySchulte@uni-muenster.de}

\abstract{Two cryogenic bath-type heat exchangers for ultra-pure noble gas applications were developed with particular emphasis on noble gas liquefaction in cryogenic distillation systems. The main objective was to construct heat exchangers for xenon from materials that do not emanate radon and that fulfill ultra-high vacuum standards. Therefore, only high-quality copper and stainless steel materials were used. Especially, large-area oxygen-free copper fins with high conductivity in a new design ensure efficient heat transfer.
One bath-type Xe-Xe heat exchanger was designed with a diameter of 50\,cm to achieve a xenon condensing capacity of at least 100\,kg/h. In order to guarantee the necessary heat transfer between the two xenon reservoirs, this heat exchanger features a specially manufactured stainless steel flange with a copper plate welded inside. We first tested our concept on a dedicated bath-type heat exchanger with a reduced diameter of 30\,cm using liquid nitrogen to liquefy the xenon. A model based on conservative assumptions such as film boiling on the nitrogen side and film condensation on the xenon side was developed and applied to caluclate the expected heat transfer for our design. We were able to demonstrate an adjustable xenon liquefaction rate of up to 113\,kg/h limited only by our measurement procedure at a cooling efficiency of $(0.98\pm 0.03)$ for the LN$_2$-Xe heat exchanger.}

\keywords{heat exchanger, radon-free, cryogenic liquids, xenon}

\arxivnumber{} 


\proceeding{N$^{\text{th}}$ Workshop on X\\
  when\\
  where}

\maketitle
\flushbottom

\section{Introduction}
Liquid noble gases are characterised by two essential properties for basic physics research \cite{aa}: Firstly, they are an excellent detector medium, providing a highly efficient scintillator and an ideal medium to drift electric charges. Secondly, liquid noble gases can be cleaned to extreme purity by using getters, distillation or other methods. Therefore, liquid noble gas-filled detectors have become important as neutrino detectors \cite{aaa,aaaa,aaaaa,aaaaaa}, in searches for dark matter \cite{a,f,g,h,hh,i,ii} and neutrinoless double beta decay \cite{b,c,d,dd,e}, and other rare event searches \cite{j} as well as for medical applications \cite{jj,jjj}. With ever-growing noble gas targets to reach better sensitivities for the aforementioned searches, increasing purity standards are needed to ensure a low background. Consequently, also the demand for noble gas liquefaction and liquid storage systems is growing. Such systems need to fulfill special requirements with respect to cleanliness and radio-purity combined with hermetic containment \cite{n}.

A particular challenge in rare event search experiments is background created by intrinsic radioactive contamination. The radioactive noble gas isotopes of radon, a daughter of the primordial decay chains of uranium and thorium, are continuously emanated from detector materials into the detection volume, e.g. by the recoil from the $^{226}$Ra progenitor-isotope decay or by diffusion. In contrast to the short-lived radon isotopes $^{219}$Rn and $^{220}$Rn,  $^{222}$Rn with a half-life of $t_{1/2} = 3.8$\,d leads to a homogeneous distribution inside the detector. The beta decays of its progenies, for instance $^{214}$Pb, can account for the dominant background in the region of interest. Thus, material screening of all components used is of crucial importance \cite{k,l,m,n}. In order to further decrease the radon-induced backgrounds, strategies for continuous removal are developed using physisorption based on van-der-Waals forces via gas chromatography \cite{o,p} or removal via cryogenic distillation \cite{q,r,s}.

The XENONnT experiment searching for Weakly Interacting Massive Particles (WIMP) features an online radon removal system based on a cryogenic distillation column. To obtain a significant radon reduction inside the liquid xenon (LXe) target of the detector, the flow towards the radon removal system has to be at least as fast as the radon decay time constant $\tau = t_{1/2}/\ln{2} = 5.5$\,d. With a xenon inventory of 8.4 tonnes, a large LXe extraction flow of 72\,kg/h through the distillation column would correspond to a radon reduction by a factor of two for radon produced inside the detector. An additional reduction by a factor of about two in XENONnT is possible by extracting the gaseous xenon (GXe) from the top of the cryostat near particular radon sources, e.g. the cables for high voltage supply and readout of the PMTs \cite{s}.

In a cryogenic distillation column for radon removal, the radon-depleted xenon can be extracted in gaseous form from the top of the column. The extracted GXe flow has to be liquefied at the output of the distillation column again before it returns to the detector. Furthermore, it has to match the LXe input flow to keep the mass balance inside the column constant. The baseline design value for the process flow is $\Dot{F}_0=72$\,kg/h as mentioned above, but flows of up to $\Dot{F}=100$\,kg/h are a desirable option for future improvements. A cooling power of $\Dot{F}_0 \cdot \Delta H^\mathrm{Xe}_{\text{vap}}=1850$\,W is required for the given flow $\Dot{F}_0$ using the xenon enthalpy for evaporation $\Delta H^\mathrm{Xe}_{\text{vap}}$ (see table \ref{tab:gas_values}) at a xenon temperature of 178\,K. Because of the large required cooling power, an energy-efficient concept is needed to re-condense the radon-depleted GXe flow to provide a LXe return. This can be realized by utilizing the reboiler at the bottom of the column as a bath-type Xe-Xe heat exchanger (HE) with a top and bottom vessel: The radon-depleted GXe in the bottom reboiler part is thermally connected with the LXe in the top reboiler part. This allows to liquefy xenon in the bottom while evaporating xenon at the top that otherwise has to be evaporated by using electrical heaters. However, this process only works if the GXe in the bottom is at a higher pressure than the LXe in the top. This will allow the GXe to condense at a higher temperature allowing a temperature gradient between the two reservoirs. Like that, the necessary heat transfer can occur. To achieve a pressure difference, the GXe is compressed using a four cylinder magnetically-coupled piston pump \cite{u}. A 50\,cm diameter HE with large-area copper between the two reservoirs fulfills both, the purity and liquefaction requirements.

In order to prepare for such a novel high flow HE system, the concept and thermodynamic stability was first tested with a smaller 30\,cm diameter prototype using a similar design. Later, this prototype functions as the top condenser in the final radon distillation system operated with liquid nitrogen (LN$_{2}$): Typically, a LXe reflux at the top of the package column is created by partially re-condensing the evaporated xenon coming from the reboiler at the column's bottom. This enhances the column's separation efficiency and lowers the probability for radon particles to escape from the column's top and to spoil the radon-depleted xenon exhaust.
The XENONnT distillation column is designed for a reflux ratio of 0.5 corresponding to a required xenon liquefaction capability of 36\,kg/h at the top condenser or an additional required cooling power of 925\,W. 

In total, two cryogenic high-purity bath-type HEs for large noble gas liquefaction flows were developed. The construction and special design aspects with the focus on the 30\,cm diameter prototype are described in section 2 along with a calculation of the achievable heat transfer. Section 3 shows the characterization of this HE with regard to the xenon liquefaction capability using LN$_{2}$ as coolant. During the tests, it was already installed as the distillation column's top condenser. In section 4, the design extension to the 50\,cm diameter Xe-Xe bath-type HE for the reboiler system is discussed followed by a conclusion.
\section{Design Aspects of the LN$_\textbf{2}$-Xe Heat Exchanger}\label{I}
Industrial available cooling systems are not readily available for our application and would have to be customized for our purposes. Even for a customized system, often the customer does not have full control over the components used and the manufacturing processes applied by the company. However, the selection of high purity material and manufacturing following standards of ultra-high vacuum components are key requirements for our system. Furthermore, we favored a system with the possibility to scale up in terms of supplied cooling power at a later stage. Additionally, we wanted to test already early the Xe-Xe HE concept using ultra clean parts. The XENONnT infrastructure features a 30 tonne LN$_{2}$ tank underground already, serving several other subsystems. For this reason, we developed the prototype 30\,cm diameter HE with LN$_{2}$ cooling, where the cooling power can be regulated by the nitrogen flow through the HE.

In order to fulfill the requirements in terms of cleanliness, radio-purity and xenon liquefaction capability several design features were included. This is discussed in the following with the focus on the prototype 30\,cm diameter HE.
\subsection{Vacuum-insulated construction and mechanical design}
\begin{figure}[htbp]
\centering 
\includegraphics[width=.8\textwidth]{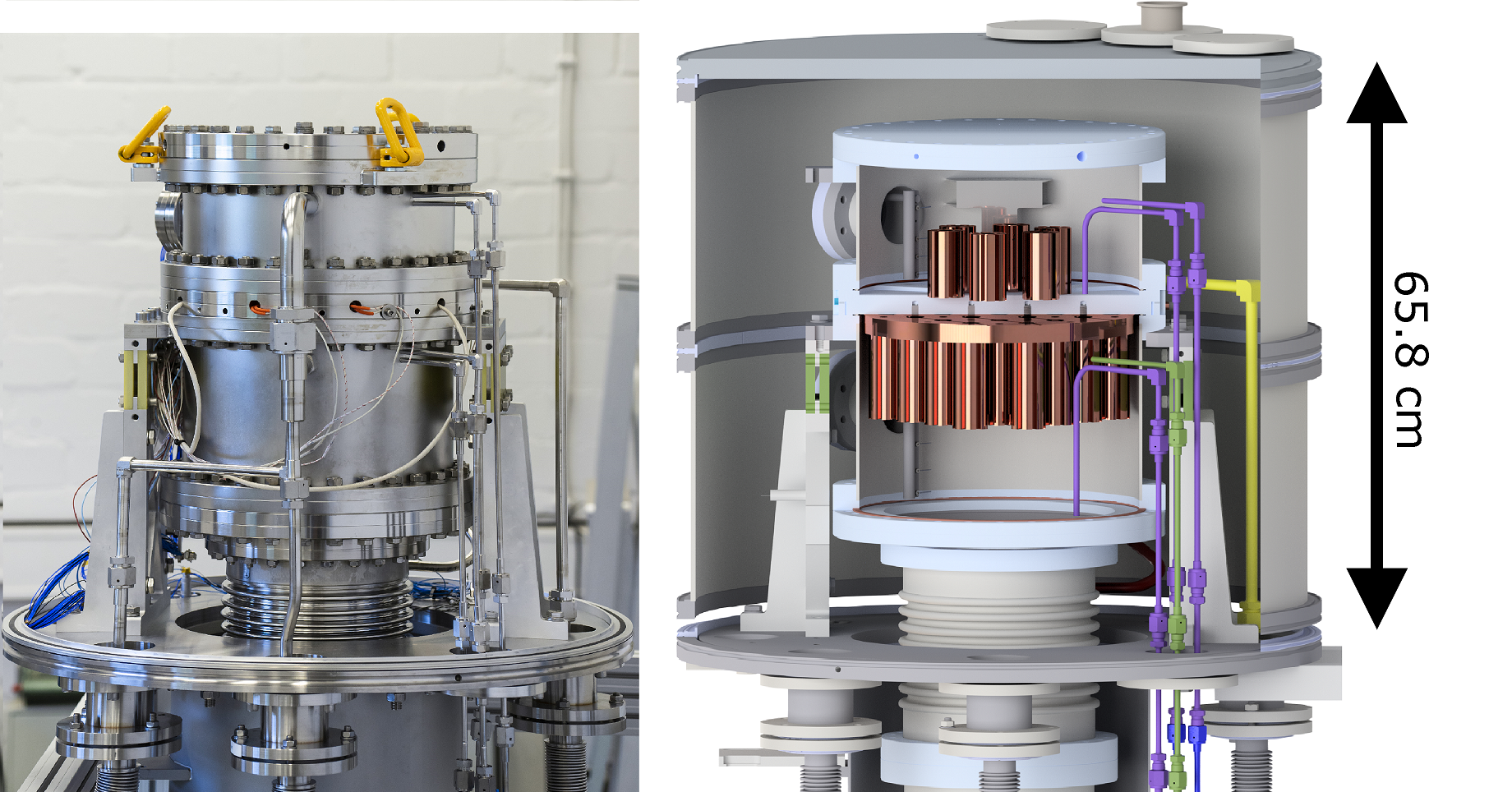}
\caption{Ultra-clean and radio-pure cryogenic bath-type heat exchanger installed as top condenser of a radon distillation column. The heat exchanger is composed of two vessels, the upper liquid nitrogen evaporation vessel and the lower xenon liquefaction vessel. Large-area oxygen-free copper fins with high conductivity provide efficient heat transfer.}
\label{fig:1} 
\end{figure}

The prototype 30\,cm diameter HE visualized in figure\,\ref{fig:1} is composed of two vessels: The top contains the cooling liquid and the bottom works as the noble gas liquefaction vessel. Beside the desired internal heat transfer between the two cryogenic fluids, reducing external heat inputs produced via thermal conduction as well as heat radiation to the environment is a key aspect of the design. In order to minimize the external heat inputs, the HE was built into an insulation housing evacuated with a turbomolecular pump combined with a rotary vane pump to a high vacuum of 10$^{-6}$\,mbar. The vessels are fixed in position within the insulation chamber via aluminum spacers. These spacers are equipped on top with preimpregnated fibre plates of Vetronite G11 to thermally decouple the inner reservoir from the outer chamber. The HE inlet and outlet pipes for xenon and nitrogen are vacuum-insulated gas lines passing the insulation chamber via bellow feedthroughs. This reduces the conduction caused heat transfer from the outer insulation chamber to the cryogenic media and avoids thermal expansion problems.

\begin{figure}[h!]
\centering 
\includegraphics[width=0.9\textwidth]{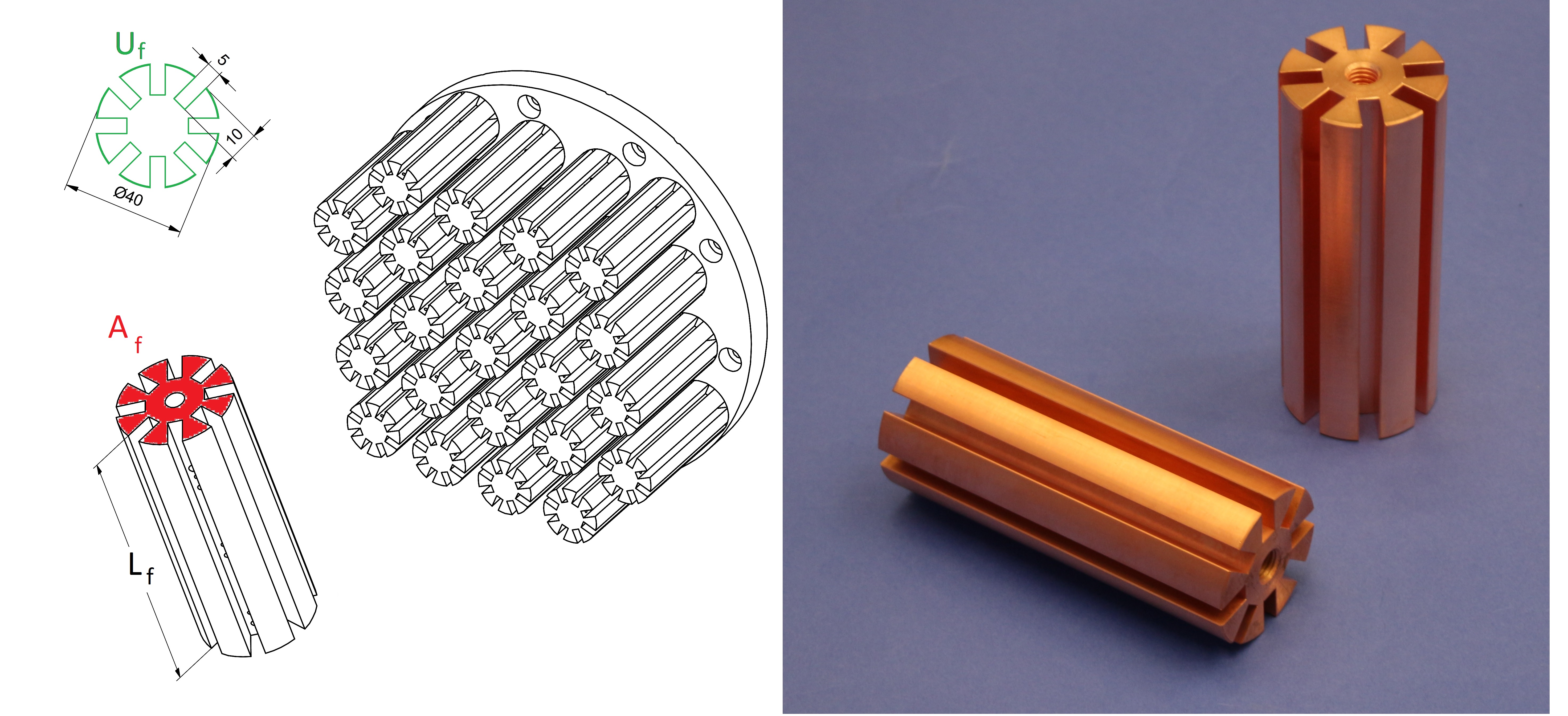}
\caption{Large-area oxygen-free high conductivity copper fins used for the heat transfer between cooling medium and operation medium within the heat exchanger. 
The cylindrical design with a diameter of 4\,cm features eight notches with a depth of 1\,cm and a width of 0.5\,cm to further enhance the perimeter to $U_\mathrm{\fins}=29$\,cm and thus the surface. The cross section of one fin is $A_\mathrm{\fins} = 8.6$\,cm$^2$. A fin with a length $L_\mathrm{\fins,GXe}=10$\,cm provides a heat transfer area of 315\,cm$^2$. For the shorter fin height $L_\mathrm{\fins,LN2}=7.5$\,cm the copper area amounts to 260\,cm$^2$.}
\label{fig:copperfins} 
\end{figure}

The top reservoir has a diameter of $D_\mathrm{ss}=30$\,cm with a height of 15\,cm. It features a 1/2" nitrogen supply line and a 1" exhaust line. Since the outlet of the vessel is positioned at a height of 10\,cm, the reservoir can be filled with 7.1\,l of LN$_2$. Inside the vessel, baffle plates at the inlet and at the outlet avoid the nitrogen to bypass the reservoir and to exit directly from the outlet without participating in the cooling process. Additionally, $N_\mathrm{\fins,LN2} = 6$ oxygen-free high conductivity copper fins shown in figure\,\ref{fig:copperfins} with a length of $L_\mathrm{\fins,LN2} = 7.5$\,cm  are installed to enlarge the surface in contact with the LN$_2$. 

The top vessel is connected via a both-sided helicoflex-type stainless steel blank flange with thickness of $d_\mathrm{ss} = 2.7$\,cm to the bottom liquefaction vessel. Reliable leak-tightness during operation is essential. It turned out, that it could only be guaranteed by special elastic recovery gaskets between both thermally coupled vessels. We use HTMS CSI gaskets with a flexible helical spring core compensating thermally caused strains of the sealing faces. A spring-surrounding copper layer ensures a leak-tight sealing to the flange face. 

Advanced screws from BUMAX of particular mechanical strength with good tensile properties (800\,MPa) are used. The special stainless-steel of the BUMAX 88 alloy allows for a larger tightening torque as well as results in a large notch impact strength at cryogenic temperatures of for example 47.4\,J at 77\,K \cite{rrr}.

The liquefaction vessel below containing the noble gas has a height of 22\,cm. A copper plate with a diameter of $D_\mathrm{Cu} = 29$\,cm and a thickness of $d_\mathrm{Cu} = 2$\,cm is directly connected to the stainless steel blank flange. At the bottom, $N_\mathrm{\fins,GXe}=23$ additional copper fins are attached to it as shown in figure\,\ref{fig:copperfins} to further enhance the total condensation surface to $A_\mathrm{tot}=0.76$\,m$^2$ and by that, to obtain a large heat transfer for the liquefaction process.
\subsection{Heat transfer for liquefaction}
\begin{figure}[h!]
\centering 
\includegraphics[width=0.9\textwidth]{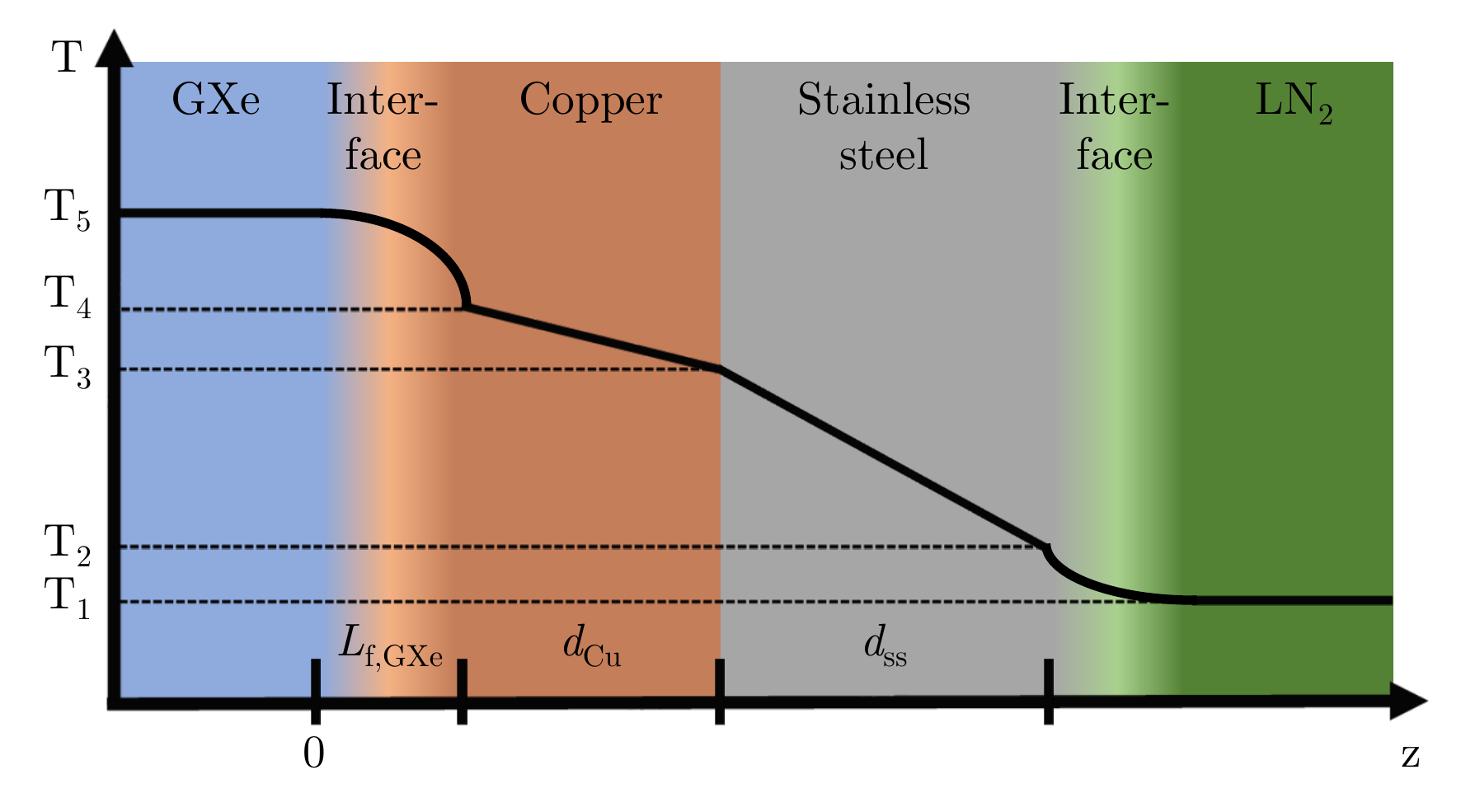}
\caption{Schematic temperature profile as a function of the z position from the GXe (left) to LN$_2$ (right) through the copper and stainless-steel plates.}
\label{fig:heattransfer}
\end{figure}

\begin{figure}[htbp]
\centering 
\includegraphics[width=0.9\textwidth]{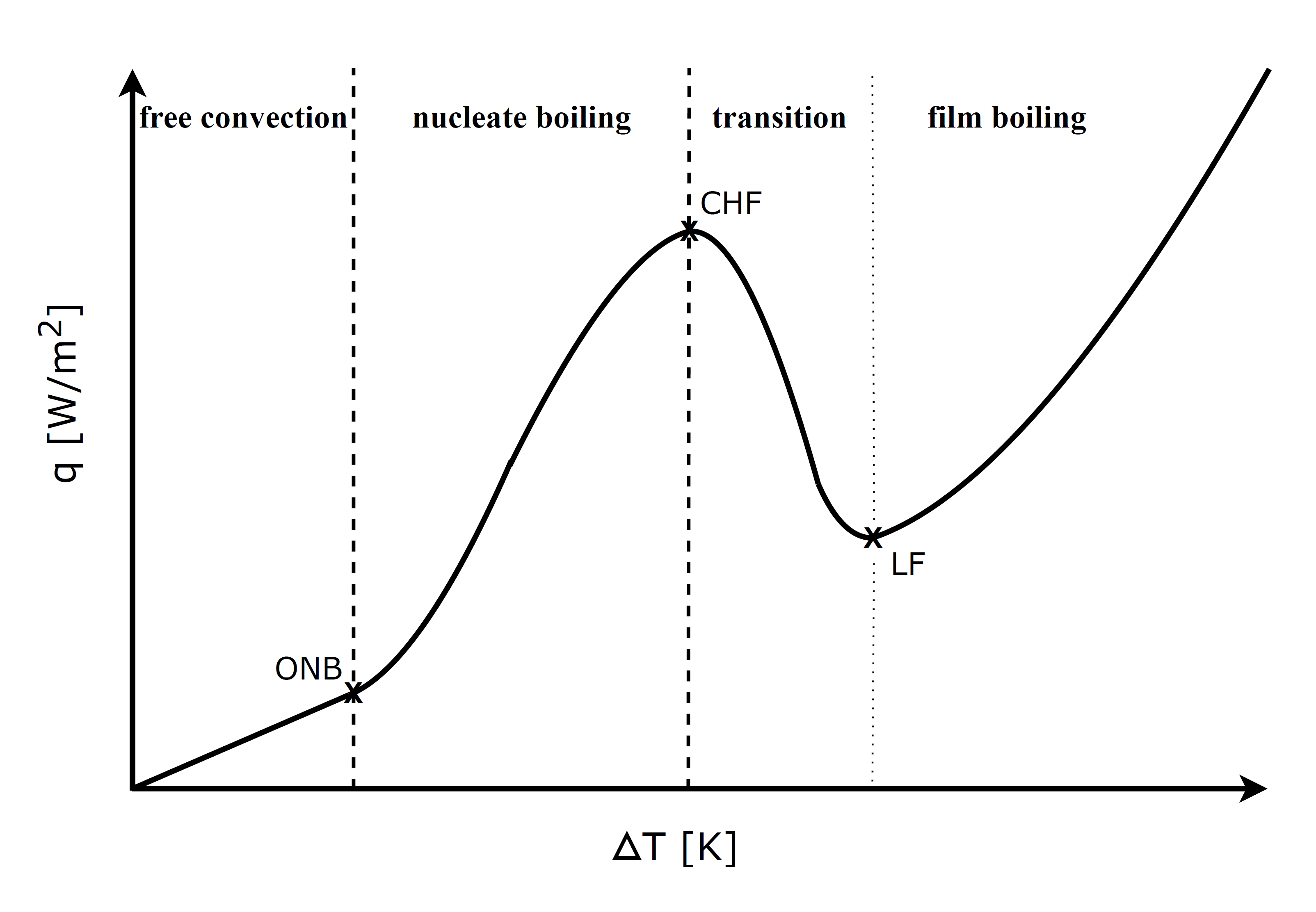}
\caption{Heat flux  by boiling of a liquid on a solid surface. The heat flux $q$ highly depends on the temperature gradient $\Delta T$ between solid and liquid leading to several heat transfer regimes: free convection, nucleate boiling, a transition section and film boiling. The transition between the regimes are called ONB (onset of nucleate boiling), CHF (critical heat flux) and LF (Leidenfrost point). Adapted from Ref. \cite{ss}.
}
\label{fig:boiling}
\end{figure}

The achievable heat transfer from the GXe in the bottom vessel of the HE to the LN$_{2}$ in the top is of particular importance for the design as it defines the ultimate HE performance. Figure\,\ref{fig:heattransfer} shows the schematical temperature profile through the different layers in z direction from the GXe to the LN$_{2}$. The heat transfer $Q$ through each layer $i$ gives rise to a temperature gradient $\Delta T_i$ depending on the corresponding heat transfer coefficient $h_i$ and the cross section of the layer $A_i$:
\begin{align} \label{film}
    Q = h_i\cdot A_i \cdot \Delta T_i.
\end{align}
The heat transfer coefficients through the stainless steel flange $h_\mathrm{ss}$ and the copper plate $h_\mathrm{Cu}$ can be calculated by using their thermal conductivity ($\lambda_\mathrm{ss}$, $\lambda_\mathrm{Cu}$) and the corresponding thicknesses ($d_\mathrm{ss}$, $d_\mathrm{Cu}$) listed in table\,\ref{tab:solid_state_values}:
\begin{align}
     h_\mathrm{ss} &= \frac{\lambda_\mathrm{ss}}{d_\mathrm{ss}} = 444\,\frac{\mathrm{W}}{\mathrm{m}^2\,\mathrm{K}} \label{eq:hss}, \\
     h_\mathrm{Cu} &= \frac{\lambda_\mathrm{Cu}}{d_\mathrm{Cu}} = 20500\,\frac{\mathrm{W}}{\mathrm{m}^2\,\mathrm{K}} \label{eq:hCu}, 
\end{align}
leading to heat transfers of
\begin{align}
  Q_\mathrm{ss} &= h_\mathrm{ss} \cdot \frac{\pi \cdot D^2_\mathrm{ss}}{4} \cdot (T_3-T_2) = 31.4 \cdot \left(\frac{T_3-T_2}{\mathrm{K}}\right)\,\mathrm{W}, \label{eq:Qss} \\
  Q_{\mathrm{Cu}} &= h_{\mathrm{Cu}} \cdot \frac{\pi \cdot D^2_\mathrm{Cu}}{4} \cdot (T_4-T_3) = 1354 \cdot  \left(\frac{T_4-T_3}{\mathrm{K}}\right)\,\mathrm{W} \label{eq:QCu}.
\end{align}
\begin{table}[h!]
    \centering
    \caption{Thermal conductivities at cryogenic temperatures \cite{rrrr} and dimensions of different stainless steel and copper components.} \label{tab:solid_state_values}
    \begin{tabular}{c|c||c|c}
         $\lambda_\mathrm{ss}$  & 12 W/(m$\cdot$K) & $D_\mathrm{ss}$ & 30\,cm\\
         	&				&	 $d_\mathrm{ss}$ & 2.7\,cm\\ \hline
	 $\lambda_\mathrm{Cu}$  & 410 W/(m$\cdot$K) & $D_\mathrm{Cu}$ & 29\,cm\\ 
         	&				&			  $d_\mathrm{Cu}$ & 2.0\,cm\\
		&				&	 $A_\mathrm{\fins}$ & 8.6\,cm$^2$\\
		&				&	$U_\mathrm{\fins}$ & 29\,cm\\
   		&				&	$L_\mathrm{\fins,LN2}$ & 7.5\,cm\\
   		&				&	$L_\mathrm{\fins,GXe}$ & 10\,cm\\
        &				&	$N_\mathrm{\fins,LN2}$ & 6\\   		
        &				&	$N_\mathrm{\fins,GXe}$ & 23\\
             \end{tabular}
\end{table}
\newpage
\paragraph{Interface between LN$_2$ and stainless steel flange:}
The heat transfer of the transition layer between the solid stainless steel surface at a temperature $T_2$ and the liquid nitrogen at a temperature $T_1 = 77$\,K (at 1\,bar) highly depends on the corresponding heat transfer regime and on the temperature gradient $\Delta T_\mathrm{LN2-ss} := T_2-T_1$:  
Figure\,\ref{fig:boiling} illustrates that the connection of the heat transfer $Q$ through an area $A$ given by the heat flux $q=Q/A$ with the temperature gradient $\Delta T$ depends on the underlying heat transfer mechanism. 
For small temperature differences $\Delta T$, the heat transfer is dominated by free convection resulting in a small heat transfer. At the onset of nucleate boiling (ONB), bubble formation occurs, where the vapor inside is heated to a temperature larger than the temperature of the saturated liquid. However, most of the solid surface is still covered by liquid. Above a critical heat flux (CHF), the heat transfer is reduced again since most of the surface is covered with large bubbles. At a critical temperature, the Leidenfrost (LF) point, a continuous insulating vapor film is created. Further increase of the solid temperature and thus the temperature difference $\Delta T$ enhances the heat flux again.

As a first conservative assumption, we expect that the temperature gradient $\Delta T_\mathrm{LN2-ss}$ 
will be large enough to be in the film boiling regime above the Leidenfrost point. According to Ref. \cite{v}, total film boiling is expected for liquid nitrogen on a stainless-steel surface at
\begin{equation}\label{eq:minDeltaTLF}
  \Delta T_\mathrm{LN2-ss}  \geq 35\,\mathrm{K}.  
\end{equation}
However, this number is only an estimate since the exact onset of total film boiling depends also on the surface roughness of the stainless-steel \cite{w}. For film boiling on a horizontal surface, hydrodynamic considerations have to be taken into account for bubble formation. According to Ref. \cite{w}, the heat transfer coefficient $h_\mathrm{LN2-ss}(\Delta T_\mathrm{LN2-ss})$ is given by
\begin{align}
\label{eq:film_boiling_calculation}
h_\mathrm{LN2-ss}(\Delta T_\mathrm{LN2-ss}) &= 0.425 \cdot \left[\frac{g\cdot \rho_{\mathrm{GN}_2} (\rho_{\mathrm{LN}_2}-\rho_{\mathrm{GN}_2}) \cdot k_{\mathrm{GN}_2}^3\cdot \Delta H^\mathrm{{N_2}}_{\text{vap}}}{\mu_{\mathrm{GN}_2}\cdot (T_2-T_1)\cdot \sqrt{\frac{\sigma_{\mathrm{LN}_2}}{g(\rho_{\mathrm{LN}_2}-\rho_{\mathrm{GN}_2})}}}\right]^\frac{1}{4} \notag \\
                                            &= 350 \cdot \left( \frac{\mathrm{K}}{T_2 - T_1} \right)^{1/4} \,\frac{\mathrm{W}}{\mathrm{m}^2\,\mathrm{K}}~,
\end{align}
with $g$ being the gravity acceleration, $\rho_{\mathrm{GN}_2}$ the vapor density, $\rho_{\mathrm{LN}_2}$ the liquid density at the boiling point, $k_{\mathrm{GN}_2}$ the vapor thermal conductivity, $\Delta H_{\text{vap}}^{\mathrm{N}_2}$ the nitrogen enthalpy of evaporation, $\mu_{\mathrm{GN}_2}$ the vapor viscosity, and $\sigma_{\mathrm{LN}_2}$ the liquid nitrogen surface tension. All values are summarized in table \ref{tab:gas_values}. Although equation \ref{eq:film_boiling_calculation} was derived for a horizontal liquid-solid interface, we add the vertical fin surfaces to the stainless steel flange
neglecting any temperature gradient over the length of the fins because of their large heat conductivity.
Altogether, this yields a total interface area of 2012\,cm$^2$ with a heat transfer of
\begin{align} \label{eq:Q_from_T2}
    Q_\mathrm{LN2-ss} &= h_\mathrm{LN2-ss}(\Delta T_\mathrm{LN2-ss}) \cdot \left(  \frac{\pi \cdot D^2_\mathrm{ss}}{4} + N_\mathrm{\fins,LN2} \cdot L_\mathrm{\fins,LN2} \cdot U_\mathrm{\fins} \right) \cdot (T_2-T_1) \notag \\
    &= 70.5 \cdot  \left(\frac{T_2-T_1}{\mathrm{K}} \right)^{3/4}\,\mathrm{W}.
\end{align}
The minimal heat transfer is calculated to be $Q_\mathrm{LN2-ss} \geq 1014$\,W considering the minimum temperature difference for film boiling (equation\,\ref{eq:minDeltaTLF}).
\begin{table}[h!]
    \centering
    \caption{Thermodynamical parameters of nitrogen at 1\,bar and 77\,K (left), xenon at 2\,bar and 178\,K (middle) and xenon at 3.5\,bar and 190\,K $^*$ (right).
    All values are from Ref. \cite{rr}. The value for $\sigma_{\mathrm{LN}_2}$ is from Ref. \cite{rs}. \label{tab:gas_values}}
    \begin{tabular}{l|c|c|c|l}
    Paramter & Nitrogen & Xenon & Xenon $^*$ & Unit\\\hline \hline
         $\rho_{\mathrm{gas}}$ & 4.56 
            & 18.69 & 31.19 & kg/m$^3$ \\
         $\rho_{\mathrm{liquid}}$ & 806.61 
            & 2854.70 & 2768.36 & kg/m$^3$ \\\hline
         $k_{\mathrm{gas}}$     & 0.0072 
            & 0.0033& 0.0036 & W/(m$\cdot$K) \\
         $k_{\mathrm{liquid}}$     &  0.1451
            & 0.0671& 0.06213 & W/(m$\cdot$K) \\\hline

         $\mu_{\mathrm{gas}}$   & $5.43\cdot 10^{-6}$ 
            & $13.77\cdot 10^{-6}$ & $14.85\cdot 10^{-6}$ & \,Pa$\cdot$s \\
         $\mu_{\mathrm{liquid}}$   & $1.6 \cdot 10^{-4}$ 
            & $4.1\cdot 10^{-4}$   & $3.4 \cdot 10^{-4}$ & \,Pa$\cdot$s \\ \hline
             $\Delta H_{\text{vap}}$ & 199.3 
             & 92.5& 89.3 & kJ/kg \\
         $\sigma$ & 8.94& 15.93 & 13.86 &mN/m\\
    \end{tabular}
\end{table}

\paragraph{Interface between copper plate with fins and GXe:}
Large-area vertical copper fins are used to increase the liquefaction surface per unit volume considering that surface condensation only appears when the temperature of the xenon $T_5= 178$\,K at a pressure of 2\,bar can be reduced below its saturation temperature. Furthermore, the fins help to compensate the fact that the condensate on the surface itself corresponds to a resistance for further heat transfer. The total cross section of the $N_\mathrm{\fins,GXe}=23$ fins amounts to $N_\mathrm{\fins,GXe} \cdot A_\mathrm{\fins} = 197$\,cm$^2$.

Although drop-wise condensation is more efficient, for our conservative estimation of the heat transfer coefficient we assume that only film condensation takes place on the xenon side. According to Ref.\,\cite{ss}, the average convection coefficient $h_\mathrm{Cu-GXe}$ for a temperature gradient $\Delta T_\mathrm{Cu-GXe} := T_5-T_4$ on a vertical surface with the length $L_\mathrm{\fins,GXe}$ is given by
\begin{align}  \label{eq:film2}                              
h_\mathrm{Cu-GXe}(\Delta T_\mathrm{Cu-GXe} ) &= 0.943 \left[\frac{g\cdot \rho_\mathrm{LXe} (\rho_\mathrm{LXe}-\rho_\mathrm{GXe}) k_\mathrm{LXe}^3\cdot \Delta H^\mathrm{Xe}_{\text{vap}}}{\mu_\mathrm{LXe}\cdot (T_5 - T_4)\cdot L_\mathrm{\fins,GXe}}\right]^{1/4} \notag \\
                                             &= 2558 \cdot \left( \frac{\mathrm{K}}{T_5-T_4} \right)^{1/4} \,\frac{\mathrm{W}}{\mathrm{m}^2\,\mathrm{K}},
\end{align}   
with $\rho_\mathrm{LXe}$ being the liquid density, $\rho_\mathrm{GXe}$ the gaseous density, $k_\mathrm{LXe}$ the liquid thermal conductivity, $\mu_\mathrm{LXe}$ the dynamic liquid viscosity, and the temperature difference between the saturated vapor $T_5$ and the copper surface $T_4$. All values are listed in table \ref{tab:gas_values}. For a conservative estimate, we consider only laminar and wave-free film condensation since turbulent flow regimes further increase the condensation rate.
\begin{figure}[htbp]
\centering 
\includegraphics[width=0.9\textwidth]{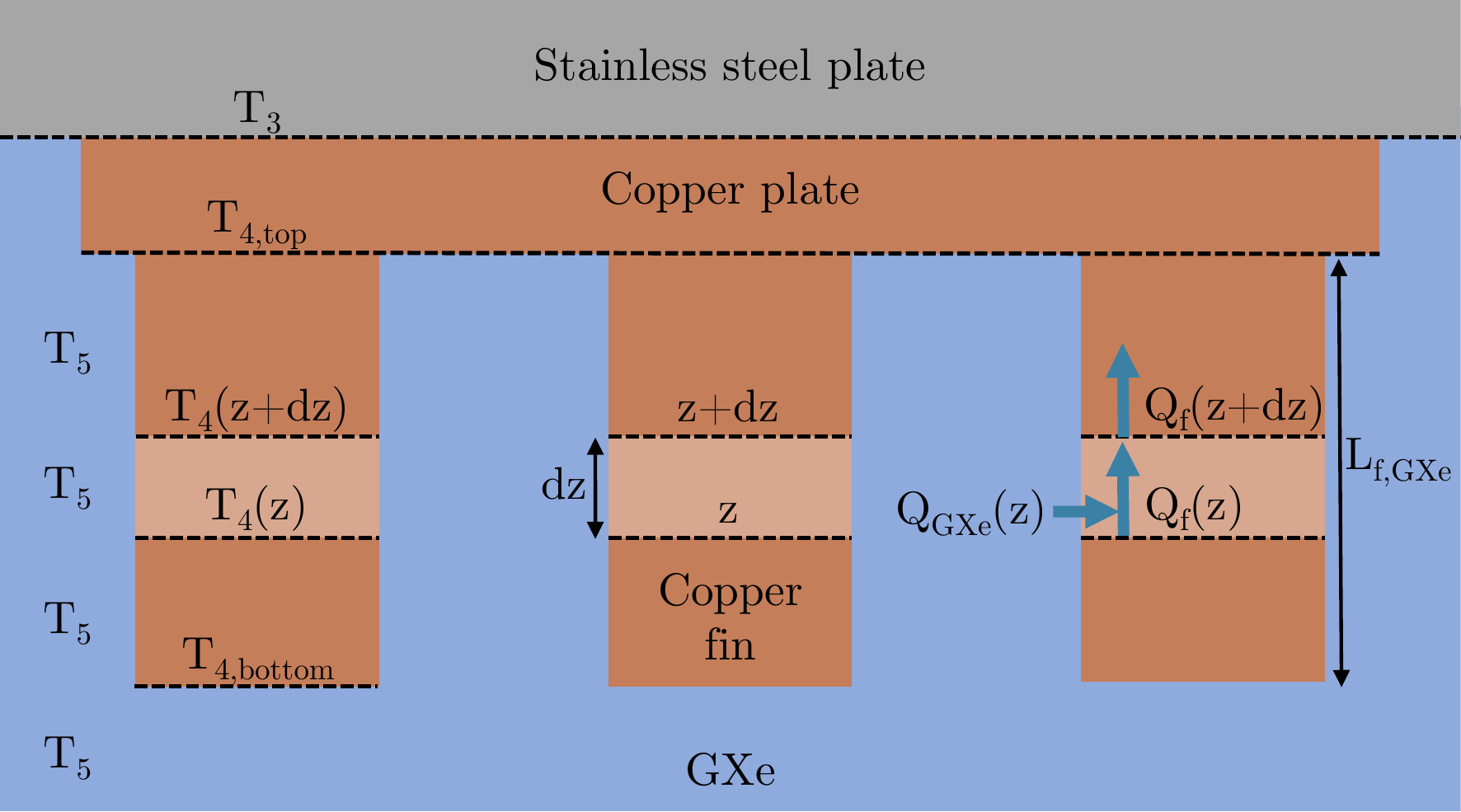}
\caption{Scheme to explain the temperature $T_4(z)$ and heat transfer $Q(z)$ profiles in the copper fins.} 
\label{fig:fins_temperature}
\end{figure}
The surface temperature $T_4$ of the copper facing the saturated xenon vapor in equation \ref{eq:film2} is not constant because the fins accumulate the heat from the vapor and transport it to the copper plate creating a temperature difference $T_\mathrm{4\,top}-T_\mathrm{4\,bottom}$ over the length of the fins as illustrated in figure \ref{fig:fins_temperature}.
Therefore, we can separate the heat transfer through the area of the copper plate not covered by fins $Q_\mathrm{no-fins}$ from the heat transfer through the fins 
$Q_\mathrm{\fins} = Q_\mathrm{Cu-GXe}-Q_\mathrm{no-fins}$. The former is given by
\begin{align} \label{eq:Qnofins}
    Q_\mathrm{no-fins} &= h_\mathrm{Cu-GXe}(T_5-T_\mathrm{4\,top}) \cdot \left(  \frac{\pi \cdot D^2_\mathrm{Cu}}{4} - N_\mathrm{\fins,GXe} \cdot A_\mathrm{\fins} \right) \cdot (T_5-T_\mathrm{4\,top})\notag \\
                        &= 118 \cdot  \left(\frac{T_5-T_\mathrm{4\,top}}{\mathrm{K}} \right)^{3/4}\,\mathrm{W}.
\end{align}
For the heat transfer through the fins, we obtain a set of coupled differential equations describing the height-dependent heat transfer $Q_\mathrm{\fins}(z)$: 
\begin{align}
    Q_\mathrm{\fins}(z)    &=  \frac{\lambda_\mathrm{Cu}}{dz} \cdot N_\mathrm{\fins,GXe} \cdot  A_\mathrm{\fins} \cdot \left( T_4(z)-T_4(z+dz) \right) \notag \\
                           &= -N_\mathrm{\fins,GXe} \cdot  A_\mathrm{\fins} \cdot \lambda_\mathrm{Cu} \cdot \frac{dT_4}{dz}(z), \label{eq:qfins} \\
    Q_\mathrm{\fins}(z+dz) &=  Q_\mathrm{\fins}(z) + Q_\mathrm{GXe}(z) \notag \\
                           &=  Q_\mathrm{\fins}(z) + dz \cdot N_\mathrm{\fins,GXe}\cdot U_\mathrm{\fins}\cdot h_\mathrm{Cu-GXe}\left(\Delta T_\mathrm{Cu-GXe}(z)\right)\cdot (T_5-T_4(z)), 
    \label{eq:qfins_dz}
\end{align}
with $\Delta T_\mathrm{Cu-GXe}(z) := T_5-T_4(z)$. Combining equations \ref{eq:qfins} and \ref{eq:qfins_dz} leads to
\begin{align}
    \frac{d Q_\mathrm{\fins}}{dz}(z) &= -N_\mathrm{\fins,GXe}\cdot  A_\mathrm{\fins} \cdot \lambda_\mathrm{Cu} \cdot \frac{d^2T_4}{dz^2}(z) \notag \\
                                    &= N_\mathrm{\fins,GXe}\cdot U_\mathrm{\fins}\cdot h_\mathrm{Cu-GXe}\left(\Delta T_\mathrm{Cu-GXe}(z)\right) \cdot (T_5-T_4(z))
\end{align}
and thus,
\begin{align}
     \frac{d^2T_4}{dz^2}(z) &=  - \frac{ U_\mathrm{\fins}}{A_\mathrm{\fins} \cdot \lambda_\mathrm{Cu}} \cdot h_\mathrm{Cu-GXe}\left(\Delta T_\mathrm{Cu-GXe}(z)\right) \cdot (T_5-T_4(z)) \notag \\
                            &= -2104 \left(\frac{T_5-T_4(z)}{\mathrm{K}} \right)^{3/4} \frac{\mathrm{K}}{\mathrm{m^2}}. \label{eq:fins_temp2}
\end{align}
Equation \ref{eq:fins_temp2} is solved  by
\begin{equation} \label{eq:T4}
    T_4(z) = - 1.99 \cdot 10^6 \cdot (z-z_0)^8 \,\frac{\mathrm{K}}{\mathrm{m^8}} +T_5,
\end{equation} 
with the integration constant $z_0$ defined by the boundary conditions. Equations \ref{eq:qfins} and \ref{eq:T4} yield a heat transfer profile over the fins of
\begin{equation} \label{eq:Qfins}
     Q_\mathrm{\fins}(z) = -N_\mathrm{\fins,GXe} \cdot  A_\mathrm{\fins} \cdot \lambda_\mathrm{Cu} \cdot \frac{dT_4}{dz}(z) 
     =  1.29 \cdot 10^8 \cdot \left( z-z_0\right)^{7} \frac{\mathrm{W}}{\mathrm{m^7}}.
\end{equation}
\begin{figure}[h!]
\centering 
\includegraphics[width=\textwidth]{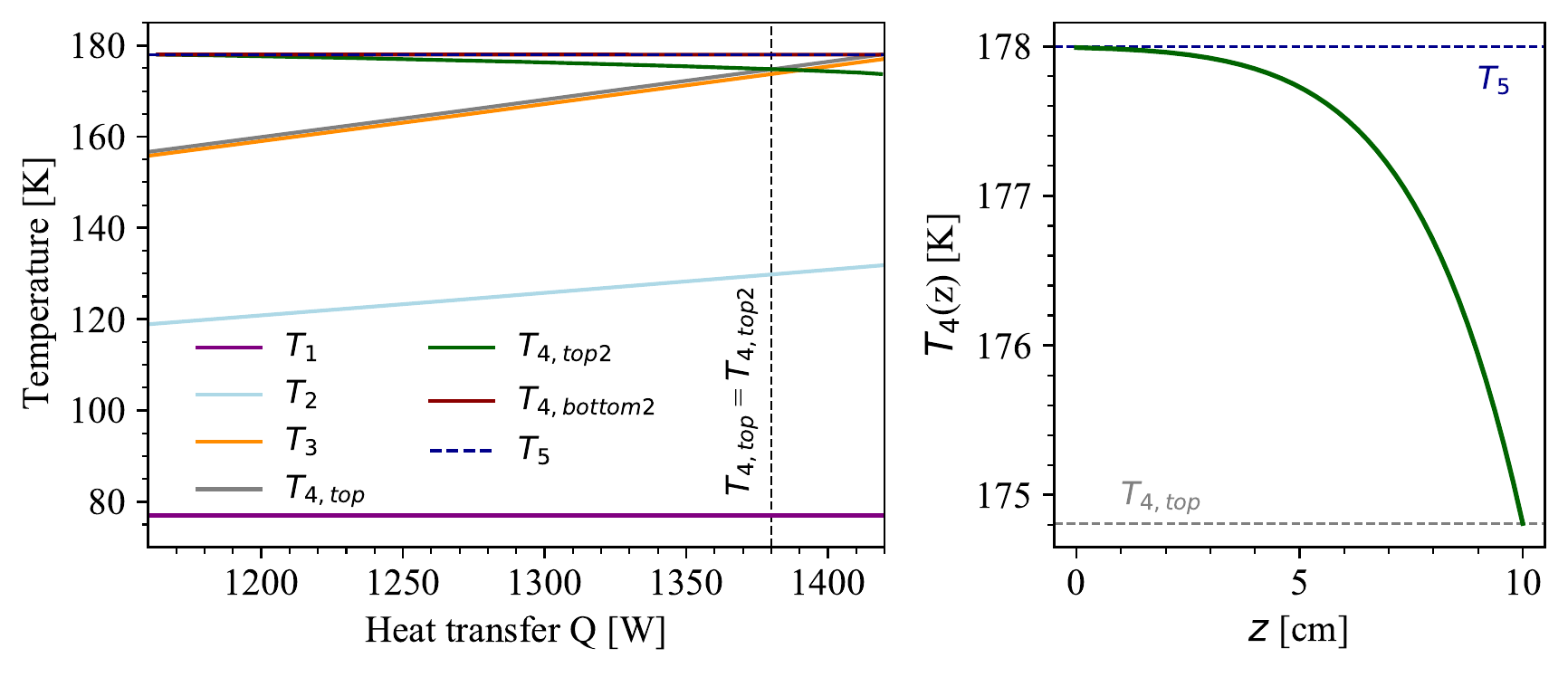}
\caption{Temperatures as function of the heat transfer $Q$ (left) and temperature profile $T_\mathrm{4}(z)$ across the fins in GXe for the matching heat transfer $Q=1380$\,W at $T_\mathrm{4,top} = T_\mathrm{4,top2}$ (right).}
\label{fig:temperatures_versus_Q} 
\end{figure}

In the following, we search for a set of temperatures that fulfills the condition
\begin{equation}
    Q (T_1, T_2, T_3, T_4, T_5) = Q_\mathrm{LN2-ss} = Q_\mathrm{ss} = Q_\mathrm{Cu} = Q_\mathrm{Cu-GXe}.
\end{equation}
Furthermore, we require that temperature $T_2$ obeys equation \ref{eq:minDeltaTLF} such that film boiling can be assumed. With that, the heat transfer $Q_\mathrm{LN2-ss}$ can be computed using equation \ref{eq:Q_from_T2} assuming a liquid nitrogen temperature of $T_1 = 77$\,K at a pressure of 1\,bar. The temperatures $T_3$ and $T_4 = T_\mathrm{4\,top}$ are then calculated using equations \ref{eq:Qss} and \ref{eq:QCu} under the assumption of $Q_\mathrm{LN2-ss} = Q_\mathrm{ss} = Q_\mathrm{Cu}$. The temperature $T_\mathrm{4,top}$ is inserted into equation \ref{eq:Qnofins} yielding the heat transfer $Q_\mathrm{\fins}$ which has to be transferred via the copper fins in GXe.

In a second step, we assume that the temperature $T_5$ is 178\,K for gaseous xenon at a pressure of 2\,bar. From here, we calculate backwards and derive a second value for $T_4 = T_\mathrm{4,top2}$ for the maximum possible heat transfer $Q_\mathrm{Cu-GXe}$ (with boundary condition $T_\mathrm{4,bottom2} \approx T_5$) by using equations \ref{eq:T4} and \ref{eq:Qfins}.

Figure \ref{fig:temperatures_versus_Q} shows that for $Q=1380$\,W the condition $T_\mathrm{4,top} = T_\mathrm{4,top2}$ is fulfilled defining the heat transfer and the corresponding set of temperatures which this HE should provide under the specified assumptions. This value for the heat transfer $Q$ is over-fulfilling the required design heat transfer of 925\,W for the top condenser by a factor of 1.5. A comparison with measurements is discussed in section 3.

In these measurements, the HE was operated with a nitrogen pressure of around 4 bar determined by the XENONnT LN$_2$ tank. This changes the temperature $T_1$ to 91\,K \cite{rr}. There is no reference in literature if film boiling is applicable at such a nitrogen pressure. However, when assuming the new $T_1$ and solving equations \ref{eq:Qss} to \ref{eq:Qfins}, a maximum heat transfer of $Q = 1356$\,W is obtained.
\subsection{Controlling and Monitoring}
\label{subsec:controlling_monitoring}
The LN$_2$ vessel contains an inlet and outlet pipe and two quarter inch lines for differential pressure measurements to determine the liquid level via a MKS Baratron 226A sensor. The capacitance manometer measures the pressure difference
\begin{align}
    \Delta p = p_x - p_r
\end{align}
between  the pressure of the gas phase in the vessel $p_\mathrm{r}$ and the pressure of the gas phase in the quarter inch tube caged by the liquid $p_x$. The difference $\Delta p$ then determines the liquid level $l$ via
\begin{align}
    l = \frac{g\cdot \rho_{\mathrm{LN}_2}}{\Delta p}.
\end{align}

Additionally, three Honeywell HEL-705 platinum Resistance Temperature Detectors (RTDs) are placed inside the LN$_2$ vessel as a conventional level meter indicating when the liquid level reaches a height of 1, 3 and 5\,cm. In the xenon vessel, additional eight RTDs are installed to monitor the temperature of the copper plate itself as well as the temperature profile across different copper fins attached to the plate. A Lakeshore DT-670 silicon diode measures the temperature in the stainless steel flange between the xenon and nitrogen vessels. These measurements are used to avoid xenon freezing and allow for a stable liquefaction process.

In order to regulate and to balance the heat transfer and to avoid xenon freeze-out, a heating system is installed inside the connector flange between both vessels: Nine homogeneously distributed WEMA S5105 315\,W heating cartridges (d = 1.27\,cm, l = 6.4\,cm) are controlled by a 3\,kW Kniel VE3PUI2 programmable power supply. This makes it possible to also compensate fluctuations of the nitrogen supply. Additionally, the nitrogen flow through the top vessel, and by that the available cooling power, can be adjusted at the outlet via a flow regulating valve (Thermomess), specifically made for cryogenic temperature applications.
\section{Characterization}\label{IV}
\begin{figure}[htbp]
\centering
\includegraphics[width=0.6\textwidth]{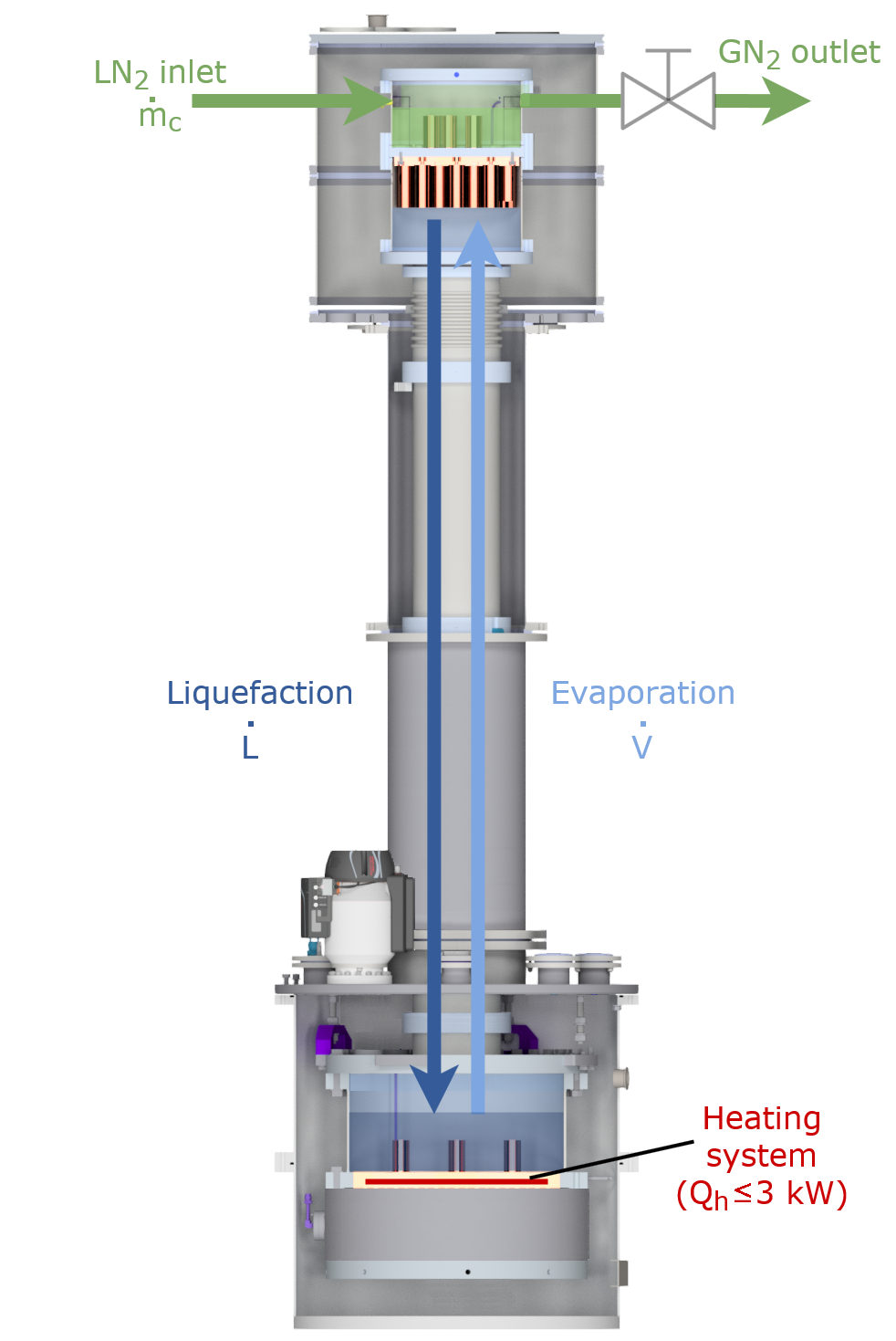}
\caption{Principle to measure the 30\,cm diameter prototype HE liquefaction capability acting as radon distillation column's top condenser: LN$_2$ is fed into the top condenser at flow rate $\Dot{m}_\mathrm{c}$ (dark green). The available cooling power $Q_\mathrm{c}$, and thus the xenon liquefaction flow $\Dot{L}$ (dark blue), can be adjusted by varying $\Dot{m}_\mathrm{c}$ via a flow regulating valve at the gaseous GN$_2$ outlet (light green). The bottom of the system contains a LXe reservoir where xenon is evaporated at a flow $\Dot{V}$ (light blue) via a PID regulated heating system with up to 3\,kW heating power. The available cooling power $Q_\mathrm{c}$ at the top can be derived from the applied heating power $Q_\mathrm{h}$ at the bottom by keeping the pressure in the system constant, neglecting other small heat losses.}
\label{fig:measurement} 
\end{figure}

One key aspect for the characterization of the 30\,cm diameter prototype HE is the xenon liquefaction capability. Several initial performance tests were carried out at the University of Münster during its development phase giving valuable input for the design concept of the 50\,cm diameter Xe-Xe HE. However, at our laboratory at Münster, the access to LN$_2$ was limited due to the usage of 100\,l dewars. Thus, only its basic functionality was tested: We were able to reach the required liquefaction flow of 36\,kg/h for the reflux at the top condenser. To measure its full potential, the final characterization of the 30\,cm diameter prototype HE happened while it was already installed as the top condenser of the full radon distillation system underground at LNGS with full access to the LN$_2$ tank. The final results are presented in this work. The basic idea to measure the liquefaction capability is shown in figure\,\ref{fig:measurement}:
LN$_2$ is fed into the top condenser at flow rate $\Dot{m}_\mathrm{c}$ (dark green). The available cooling power $Q_\mathrm{c}$, and thus the xenon liquefaction flow $\Dot{L}$ (dark blue), can be adjusted by varying $\Dot{m}_\mathrm{c}$ via a flow regulating valve at the $\mathrm{GN}_2$ outlet (light green). The bottom of the system contains a LXe reservoir where xenon is evaporated at a flow $\Dot{V}$ (light blue) via a PID regulated heating system with up to 3\,kW heating power. The available cooling power $Q_\mathrm{c}$ at the top can be derived from the applied heating power $Q_\mathrm{h}$ at the bottom by keeping the pressure in the system constant. For this purpose, the column was filled once with 30\,kg of xenon and was operated in steady state at a xenon pressure of 2\,bar.

Using the enthalpy of evaporation $\Delta H_{\text{vap}}^{\mathrm{N}_2} = 178.4$\,kJ/kg \cite{rr} of liquid nitrogen at 91\,K, the mass flow can be converted into an available cooling power $Q_\mathrm{c}$ of
\begin{align}
    Q_\mathrm{c} = \Dot{m}_\mathrm{c} \cdot \Delta H_{\text{vap}}^{\text{N}_2}\label{eq.3.1}.
\end{align}

\noindent Here, only the cooling power from the phase change of the nitrogen is taken into account.
Depending on the efficiency $\epsilon_\mathrm{c}$ of the top condenser, a cooling power $Q_\mathrm{L} = \epsilon_\mathrm{c}\cdot Q_\mathrm{c}$ is available to condense xenon at a flow rate $\Dot{L}$ of
\begin{align}
\label{eq.3.3}
    \Dot{L} = \frac{Q_\mathrm{L}}{\Delta H_{\text{vap}}^{\text{Xe}}},
\end{align}
where $\Delta H^\mathrm{Xe}_{\text{vap}} = 92.5$\,kJ/kg is the xenon enthalpy of evaporation at the given xenon temperature of 178\,K. On the other hand, depending on the efficiency of the heat transfer $\epsilon_\mathrm{h}$ in the reboiler, the heating power $Q_\mathrm{V} = \epsilon_\mathrm{h} \cdot Q_\mathrm{h}$ is available for the evaporation of xenon at a flow rate $\Dot{V}$ of
\begin{align}
    \Dot{V} = \frac{Q_\mathrm{V}}{\Delta
    H_{\text{vap}}^{\text{Xe}}}\label{eq.3.5}.
\end{align}
Here, we assume a heat transfer efficiency in the reboiler $\epsilon_h=1$ due to the good thermal conductivity of the copper in which the heaters are implemented.
At a constant xenon pressure in the system achieved by the PID regulated heating at the bottom, an equilibrium between evaporation flow $\Dot{V}$ and liquefaction flow $\Dot{L}$ is established resulting in
\begin{align}
    \Dot{V} = \Dot{L}.
\end{align}
Using equation\,\ref{eq.3.3} and \ref{eq.3.5} leads to
\begin{align}
    Q_\mathrm{V} = Q_\mathrm{L}.
\end{align}
With that, the heating power $Q_\mathrm{h}$ can be related to the cooling power $Q_\mathrm{c}$ as well as to the nitrogen mass flow $\Dot{m}_\mathrm{c}$ via
\begin{align}
   Q_\mathrm{h} = Q_\mathrm{L} =  \epsilon_\mathrm{c}\cdot Q_\mathrm{c} =  \epsilon_\mathrm{c} \cdot \Delta H_{\text{vap}}^{\text{N}_2} \cdot  \Dot{m}_\mathrm{c} =  \epsilon_\mathrm{c} \cdot \Delta H_{\text{vap}}^{\text{N}_2} \cdot (\Dot{m}_\mathrm{tot} - \Dot{m}_\mathrm{0}).
   \label{eq.3.6}
\end{align}
The total XENONnT LN$_2$ consumption rate is given by $\Dot{m}_\mathrm{tot}$ including the top condenser during the measurement that can be inferred from the XENONnT slow control data, while $\Dot{m}_\mathrm{0}$ is the consumption rate of XENONnT without the top condenser.

\begin{figure}[h!]
\centering
\includegraphics[width=0.9\textwidth]{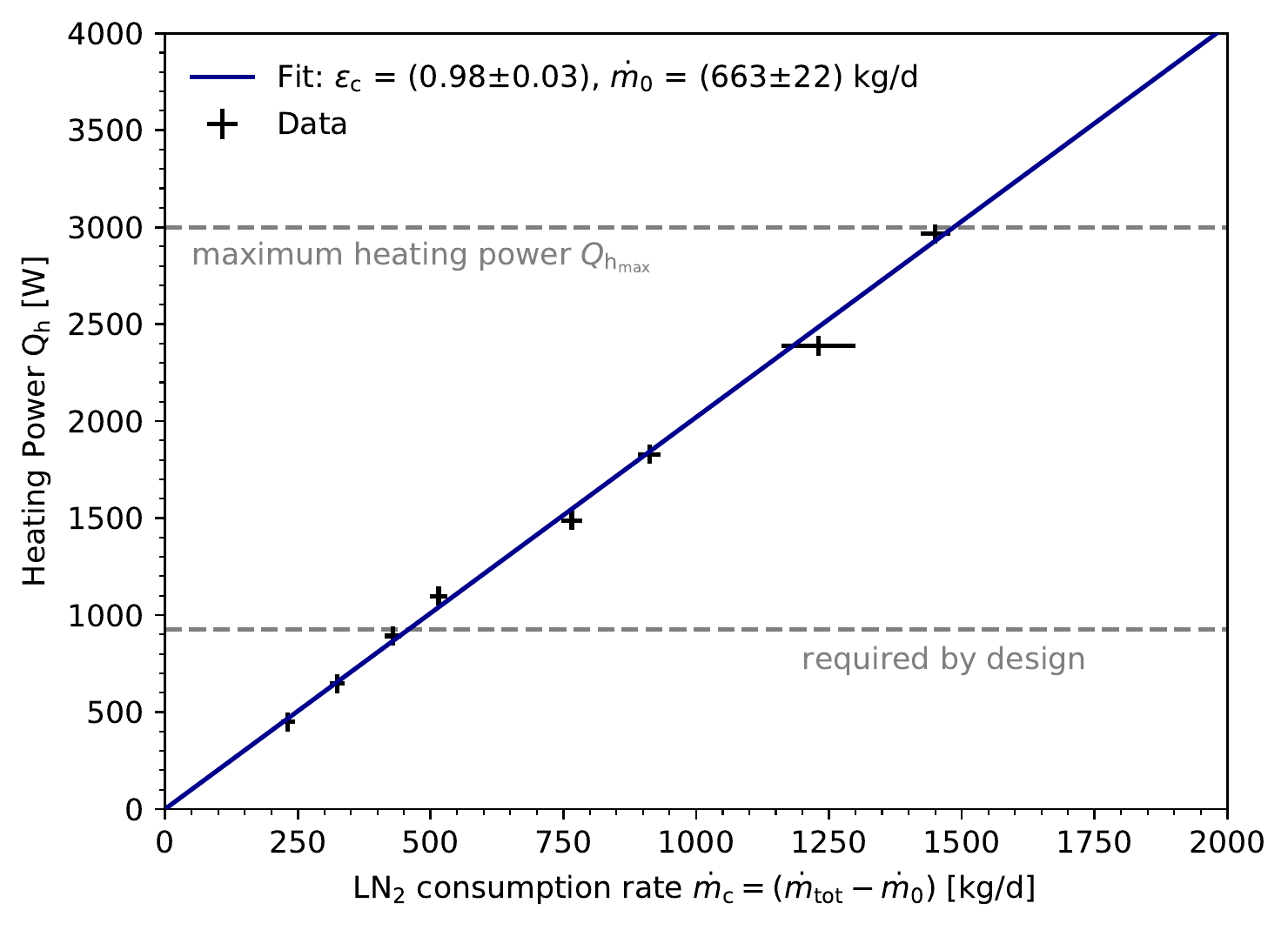}
\caption{Applied heating power $Q_\mathrm{h}$ at the reboiler as a function of the liquid nitrogen cooling consumption $\Dot{m}_\mathrm{c}$. Each point shows the equilibrium state at 2\,bar xenon pressure, such that the evaporation flow $\Dot{V}$ corresponds to the liquefaction flow $\Dot{L}$.}
\label{fig:topcondenser_performance}
\end{figure}

Figure\,\ref{fig:topcondenser_performance} shows the applied heating power $Q_\text{h}$ as a function of the LN$_2$ consumption rate $\Dot{m}_\mathrm{c}$ in black. As indicated in the figure, the measurement was limited to the maximum available heating power from the power supply at the bottom of $Q_{\mathrm{h}_\mathrm{max}} = 3$\,kW. Further increase of the cooling power was possible, but not enough heating power from the bottom was available to balance the flow rate\CW{s} $\Dot{V}$ and $\Dot{L}$. In addition, the required cooling power $Q_\mathrm{L}$ for the design reflux flow $\Dot{L}$ is highlighted: A xenon liquefaction flow $\Dot{L}$ of 36\,kg/h requires a cooling power of $Q_\mathrm{L} = 925$\,W following equation\,\ref{eq.3.3}. A linear fit based on the ride hand side of equation\,\ref{eq.3.6} to the data points yields an efficiency of $\epsilon_{\mathrm{c}} = 0.98 \pm 0.03$. Furthermore, the fit gives a consumption rate of XENONnT without the top condenser of  $\Dot{m}_\mathrm{0} = 663 \pm 22$\, kg/d. This number is consistent with measurements from a period before the top condenser measurements at standard XENONnT operation mode. 
The maximum available xenon liquefaction capability following equation\,\ref{eq.3.3} can be derived using $Q_{\mathrm{h}_\mathrm{max}} = 3$\,kW to be
\begin{align}
\label{eq:3.11}
    \Dot{L} \geq 113\,\mathrm{kg}/\mathrm{h}, 
\end{align}
and demonstrates that the nitrogen cooling system of the top condenser can provide cooling powers of
\begin{align}
\label{eq:measured_Q}
    Q_\mathrm{c} \geq 3\,\mathrm{kW}.
\end{align}
The silicon diode in the middle of the stainless steel flange between the nitrogen and xenon reservoir allows to measure a temperature profile during operation, which can be compared to the previous estimations of the heat transfer. At the above described measurement with a nitrogen pressure of 4\,bar corresponding to 91\,K and an applied heating power $Q_\mathrm{h}= 917$\,W, as an example near the design value, we measured a temperature of $T_{23}=170\,$K. Assuming a punctual temperature measurement by the silicon diode in perfect contact with the stainless steel flange, $T_2$ can be calculated using equation\,\ref{film} to be
\begin{align}
     T_2 = T_{23} - Q_h \frac{d_\mathrm{ss}/2}{\lambda_\mathrm{ss}}\cdot \frac{4}{\pi\cdot D^2_\mathrm{ss}}= 156\,\mathrm{K}.
\end{align}
This yields a temperature gradient
\begin{align}
    \Delta T_\mathrm{LN2-ss} = T_2- T_1 = 64.7\,\mathrm{K},
\end{align}
validating the assumption of total film boiling (equation\,\ref{eq:minDeltaTLF}). 

However, the achieved heat transfer $Q_\mathrm{c} \geq 3\,\mathrm{kW}$ is larger than the estimated one of 1.4\,kW. This can be explained by the conservative estimation of film condensation. It seems that the condensation mechanism is dominated by drop-wise condensation, which is expected to increase the heat transfer by one order of magnitude \cite{ss}. Additionally, the estimation of total film boiling 
(near the Leidenfrost point) gives a lower limit for the heat transfer and was based on measurements from Ref.\,\cite{v} without considerations on nitrogen pressure changes and surface properties of the stainless-steel flange. According to figure\,\ref{fig:boiling}, the transition region between nucleate and total film boiling gives much larger heat fluxes compared to total film boiling. This is due to the fact that the insulating vapor film in the nitrogen case has a two orders of magnitude lower thermal conductivity compared to the liquid.
 
Finally, the measurements show that the heat transfer mechanisms from the GXe to the copper solid is much more efficient than estimated. This findings are essential for the design extension to the larger 50\,cm Xe-Xe HE.
\section{Design Aspects of the Xe-Xe Heat Exchanger}
\label{V}
A second cryogenic HE was built to liquefy xenon with xenon. This HE acts as the distillation column's reboiler where radon-depleted and pressurized GXe is in thermal contact with the radon-enriched LXe storage that needs to be partially evaporated. In contrast to the top condenser HE, the temperature difference for the liquefaction of GXe with LXe is much smaller compared to the liquefaction of GXe with LN$_2$. This needs to be taken into account for the reboiler design.

\begin{figure}[htbp]
\centering 
\includegraphics[width=.8\textwidth]{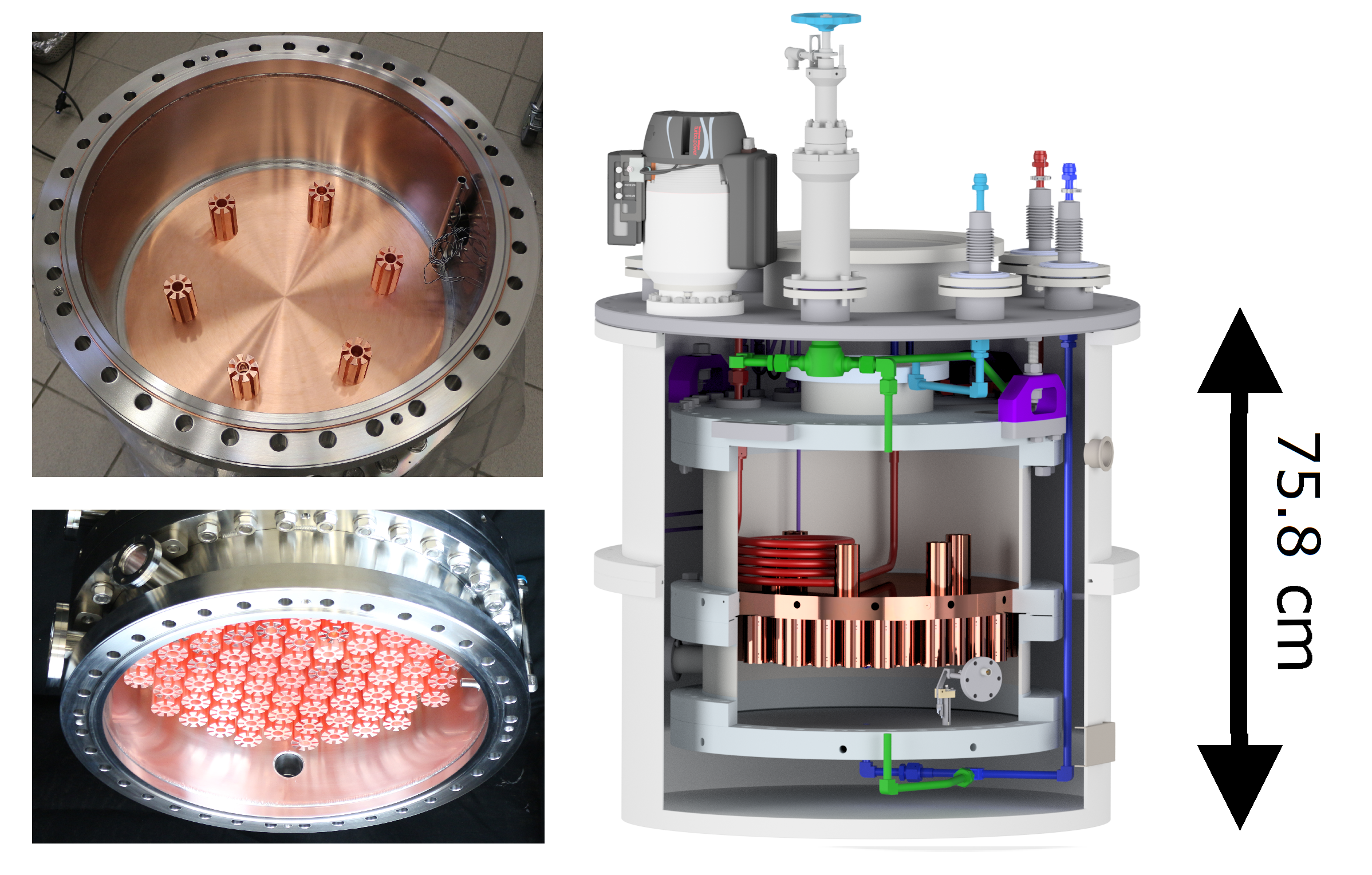}
\caption{Bath-type Xe-Xe heat exchanger acting as the distillation column's reboiler: The top vessel (left, top) stores radon-enriched LXe until the radon decays and is used as a cooling reservoir. The bottom reservoir (left, bottom) re-condenses radon-depleted GXe on large-area copper fins and is used to evaporate the LXe in the top. A copper plate was electron-welded into the bottom flange of the top vessel to guarantee absolute leak tightness between both vessels and ultra high purity to avoid any radon spoiling the radon-depleted xenon.}
\label{fig:reboiler}
\end{figure}

The reboiler is illustrated in figure\,\ref{fig:reboiler} and consists of two vessels with a diameter of \linebreak $D_\mathrm{ss,Reb}=50$\,cm. The top vessel (figure\,\ref{fig:reboiler}, left, top) is used to store LXe and trap the radon until it decays. Therefore, this is the location with the highest radon concentration in the system. The reservoir can host up to 46.5\,l LXe with a maximum filling height of 23.7\,cm. The LXe level is monitored via the hydrostatic pressure as well as by four RTDs in a similar way as explained in section \ref{subsec:controlling_monitoring}. Additionally, $N_\mathrm{\fins,top}=6$ oxygen-free high conductivity copper fins with a length of $L_\mathrm{\fins,Reb}=7.5$\,cm, similar to the ones of the top condenser shown in figure\,\ref{fig:copperfins}, are installed. Furthermore, a stainless steel spiral is immersed in the LXe to cool down the radon-depleted GXe, that is pressurized by the compressor, to its saturation temperature before it enters the bottom vessel.

The bottom vessel (figure\,\ref{fig:reboiler}, left, bottom) is then used to liquefy the radon-depleted GXe. Therefore, this is a location with low radon concentration. The vessel features a height of 20\,cm and is equipped with $N_\mathrm{\fins,bot}=79$ copper fins of $L_\mathrm{\fins,Reb}$ as well. This results in a total heat transfer surface of $A_\mathrm{tot,bot}=2.2$\,m$^2$. The maximum filling height of 10\,cm limited by the length of the copper fins results in a maximum LXe volume of 19.6\,l. The LXe level is measured with the same principles as in the top vessel. The GXe phase pressure is measured with an additional absolute pressure sensor to get insight into the liquefaction process. Additionally, a silicon diode is installed into the bottom flange to monitor the temperature.

In order to reduce the temperature gradient between the two reservoirs, a copper plate with a diameter of $D_\mathrm{Cu,Reb}=45$\,cm and a thickness of $d_\mathrm{Cu,Reb}=3.5$\,cm was electron-welded into the bottom flange of the top vessel. This method guarantees absolute leak tightness between both vessels and ultra high purity to avoid any radon spoiling the radon-depleted xenon. Two silicon diodes are installed into the copper plate to monitor the temperature and ensure stable operation. The dimensions of the different reboiler components are summarized in table\,\ref{tab:solid_state_values_reb}.

For the distillation column's designed operation, a process flow and thus, a liquefaction rate of 72\,kg/h is required corresponding to a cooling power at the bottom reboiler of $Q_\mathrm{Cu,bot}=1850$\,W. Because of the reflux coming from the top condenser in addition to the liquid feed flow, a design evaporation rate of 108\,kg/h is needed at the top reboiler corresponding to a total heat transfer of $Q_\mathrm{Cu,top}=2775$\,W. Therefore, heater cartridges are installed directly into the middle of the copper plate to add the electrically created heat $Q_\mathrm{H}$. It follows

\begin{equation}
    Q_\mathrm{Cu,top} = Q_\mathrm{Cu,bot} + Q_\mathrm{H}.
\end{equation}
More details about the process can be found in Ref. \cite{s}. 
\begin{table}[h!]
    \centering
    \caption{Dimensions of stainless steel and copper components of the reboiler.} \label{tab:solid_state_values_reb}
    \begin{tabular}{c|c}
            $D_\mathrm{ss,Reb}$ & 50\,cm\\
        	$d_\mathrm{ss,Reb}$ & 3.5\,cm\\ \hline
            $D_\mathrm{Cu,Reb}$ & 45\,cm\\ 
			$d_\mathrm{Cu,Reb}$ & 3.5\,cm\\
   			$L_\mathrm{\fins,Reb}$ & 7.5\,cm\\
   			$N_\mathrm{\fins,top}$ & 6\\
        	$N_\mathrm{\fins,bot}$ & 79\\
             \end{tabular}
\end{table}

\begin{figure}[htbp]
\centering 
\includegraphics[width=0.9\textwidth]{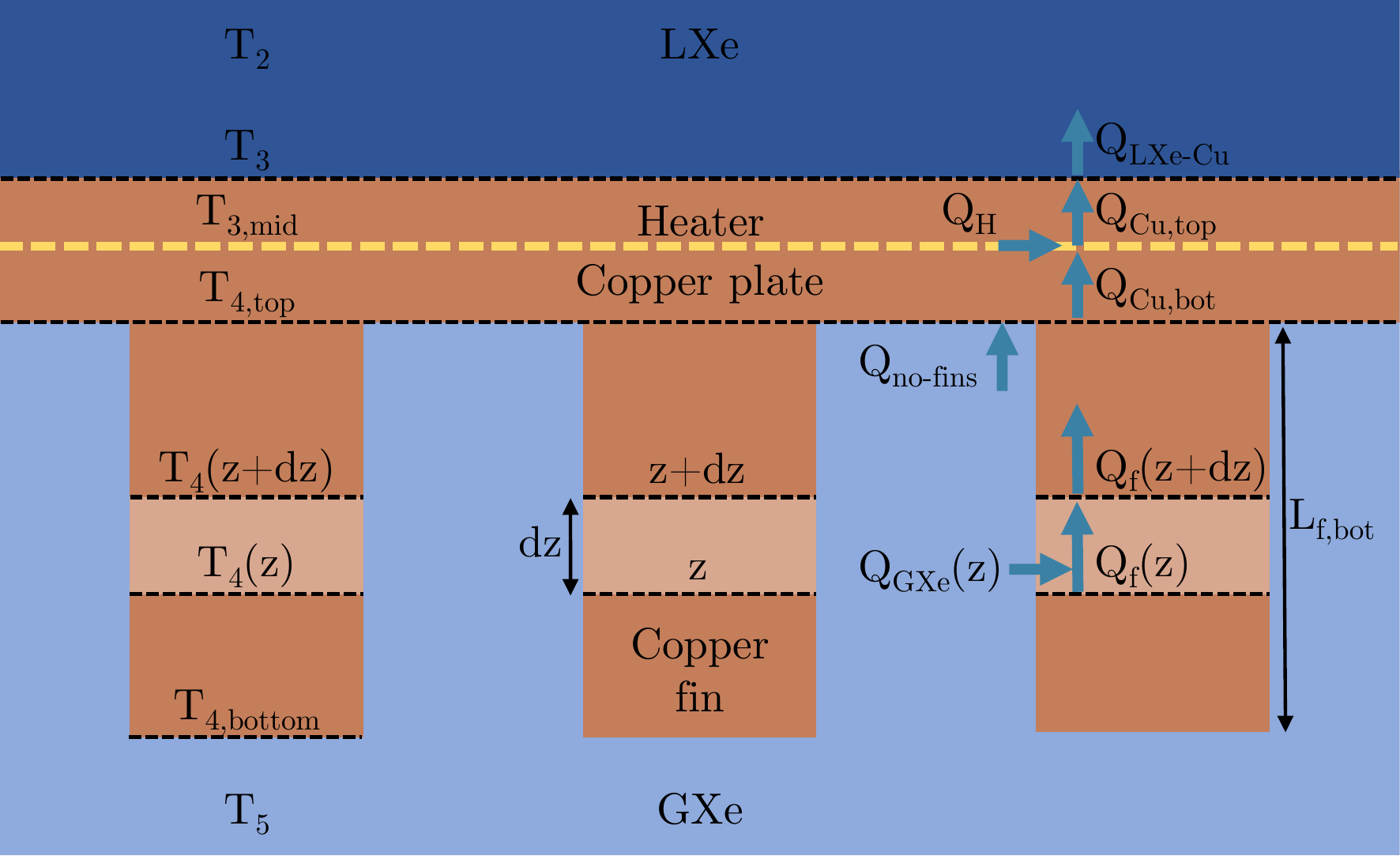}
\caption{Scheme to explain the heat transfer across the reboiler from GXe to LXe including involved temperatures.}
\label{fig:fins_temperature_reboiler}
\end{figure}

In the following, we can calculate the expected heat transfer across the reboiler analog to the calculations in section\,\ref{I} with a few modifications and assumptions. Figure \ref{fig:fins_temperature_reboiler} shows a simplified sketch of the reboiler with the involved surfaces, materials and heat transfers. We neglect the small surface covered by the stainless steel flange surrounding the copper plate as its thermal conductivity is negligible compared to the copper one. We further neglect the heat transfer involving the spiral and assume a saturated GXe in the bottom as well as a saturated LXe in the top also neglecting the small sub-cooling effect induced by hydrostatic pressure.

Due to the additional electrical heating inside the copper plate, the temperature difference between the copper plate's middle to the bottom fins is expected to be lowered yielding a reduced heat transfer. However, the temperature difference between the copper plate's middle and the LXe is expected to be larger, enhancing the heat transfer in this location. Therefore, we assume a pool nucleate boiling regime between the top surface of the copper plate and the LXe in the top reboiler vessel. There is no literature available for our specific case. In Ref.\,\cite{y}, the onset of nucleate boiling (ONB) was found for temperature differences between 3.8\,K to 18\,K for a thin platinum wire immersed in LXe at a saturation pressure of 1\,bar. In Ref.\,\cite{yy}, a copper plated resistor of a few mm$^2$ was also fully immersed in LXe at saturation pressures of about 1\,bar and 1.3\,bar. The ONB was found at around $\Delta T_\mathrm{1.0bar} = 16.9$\,K and $\Delta T_\mathrm{1.3bar} = 19.2$\,K, respectively. However, the main difference is that the bulk LXe is sub-cooled and can flow below the copper surface. In our case, the cold liquid rinsing down can only go to the copper plate surface leading to an enhanced temperature difference again and thus, leading to a better heat transfer coefficient. 

The heat transfer coefficient $h_\mathrm{LXe-Cu}$ for pool nucleate boiling of LXe on copper surface can be calculated using the Rohsenow correlation \cite{yyy}:
\begin{align}
\label{eq:htc_lxe_cu}
   h_\mathrm{LXe-Cu} \left( \Delta T_\mathrm{LXe-Cu}\right) &= \mu_\mathrm{LXe} \cdot \Delta H^\mathrm{Xe}_{\text{vap}} \cdot \left( \frac{g \cdot (\rho_\mathrm{LXe} - \rho_\mathrm{GXe})}{\sigma_\mathrm{LXe}} \right)^{1/2} \cdot \left(\frac{c_\mathrm{p,LXe}}{C_\mathrm{s,f} \cdot \Delta H^\mathrm{Xe}_{\text{vap}} \cdot Pr^{n}_\mathrm{LXe}}\right)^3 \cdot (T_3 - T_2)^2 \notag \\
   &= 29\, \frac{\mathrm{W}}{\mathrm{m^2\,K}} \cdot \left(\frac{(T_3 - T_2)}{\mathrm{K}}\right)^2,
\end{align}
with $ \Delta T_\mathrm{LXe-Cu} :=  T_3 - T_2$. Furthermore, $C_\mathrm{s,f} = 0.013$ is defined as the liquid-solid friction coefficient, and $n = 1.7$ is an exponent on the Prandtl number $Pr_\mathrm{LXe}$. Both parameters are usually experimentally determined and are not available in literature for xenon. However, for fluids other than water, commonly $n = 1.7$ is applied \cite{yyy}. For the friction coefficient, we applied the value for water on polished copper which is among the highest values available in literature \cite{yyy} leading to a lower and more conservative heat transfer coefficient according to equation \ref{eq:htc_lxe_cu}. Since the LXe surface tension and viscosity are both smaller compared to water, the LXe value for $C_\mathrm{s,f}$ on copper is expected to be smaller than the one for water. The xenon properties are calculated using the numbers from table \ref{tab:gas_values} for xenon at 178\,K. The Prandtl number is given by
\begin{equation}
    Pr_\mathrm{LXe} = \frac{c_\mathrm{p,LXe} \cdot\mu_\mathrm{LXe}}{k_\mathrm{LXe} } = 2.1 \hspace{5mm} \text{ \cite{yyy}}.
\end{equation}
In this calculation, we ignore the additional copper fins in the top reboiler for simplicity and to be conservative. The heat transfer between copper plate and LXe is then given by
\begin{equation}
  Q_{\mathrm{LXe-Cu}} = h_\mathrm{LXe-Cu} \left( \Delta T_\mathrm{LXe-Cu}\right) \cdot A_\mathrm{Cu,Reb} \cdot \Delta T_\mathrm{LXe-Cu} = 4.6\,\mathrm{W} \cdot  \left(\frac{(T_3 - T_2)}{\mathrm{K}}\right)^3
\end{equation}
with $A_\mathrm{Cu,Reb}$ being the area of the copper plate.

\begin{figure}[h!]
\centering 
\includegraphics[width=\textwidth]{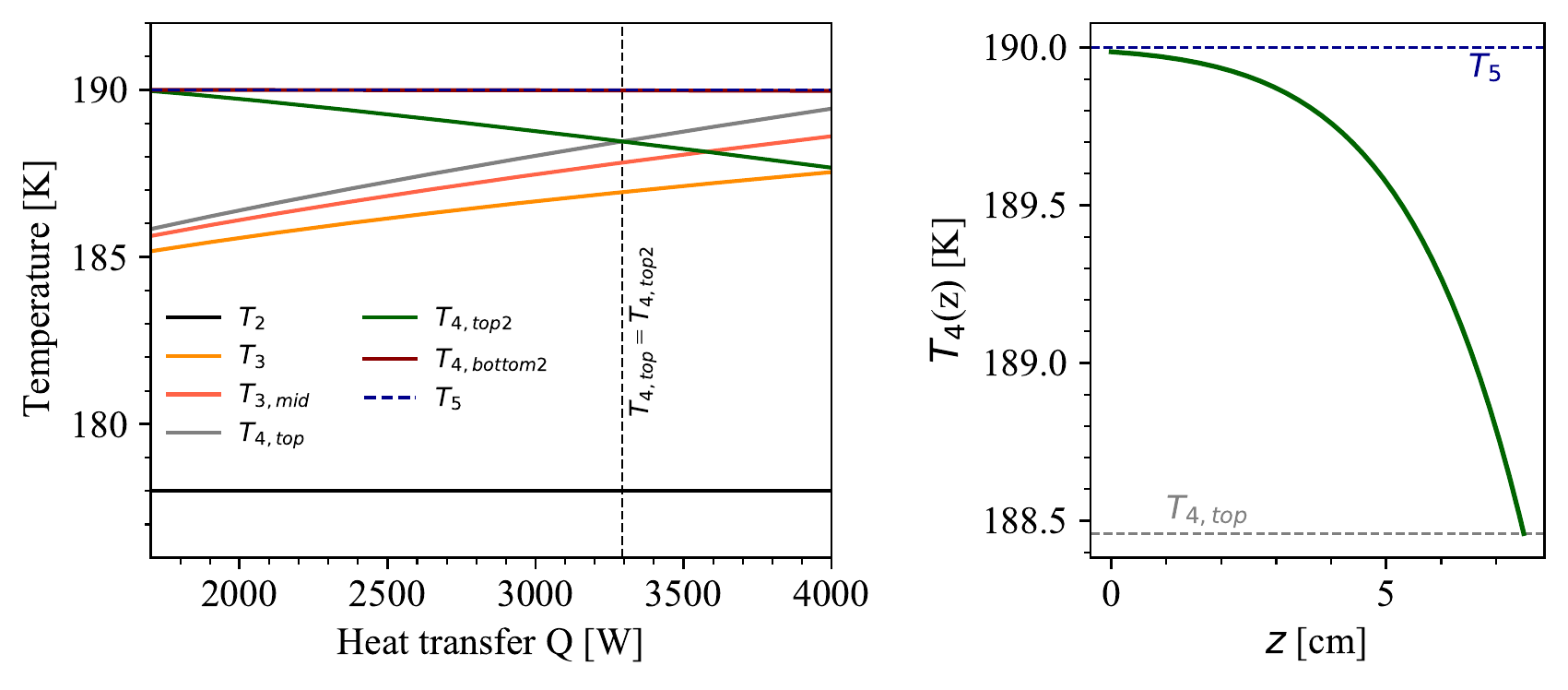}
\caption{Temperatures as function of the heat transfer $Q$ (left) and temperature profile $T_\mathrm{4}(z)$ across the fins in GXe in the reboiler (right). The matching heat transfer at $T_\mathrm{4,top} = T_\mathrm{4,top2}$ for the design pressure of 3.5\,bar in the bottom reboiler part was found to be $Q = 3293\,W$.}
\label{fig:reboiler_design_profile} 
\end{figure}

The heater position inside the copper plate is at $d_\mathrm{Cu,Reb}/2$. Therefore, we split the copper plate in two sections assuming infinitesimal thin homogeneously distributed heat input. Following equation \ref{eq:hCu}, the heat transfer coefficients for the top and bottom section are given by:
\begin{equation}
    h_\mathrm{Cu,top} = h_\mathrm{Cu,bot} = 23428 \,\frac{\mathrm{W}}{\mathrm{m}^2\,\mathrm{K}}.
\end{equation}
The heat transfers $Q_\mathrm{Cu,top}$ above the heater location and $Q_{\mathrm{Cu,bot}}$ below are
\begin{align}\label{eq:QCu_reb_top}
  Q_{\mathrm{Cu,top}} &= 3726\,\mathrm{W}  \cdot \left(\frac{T_\mathrm{3,mid}-T_3}{\mathrm{K}}\right) + Q_{\mathrm{H}}, \\
  Q_{\mathrm{Cu,bot}} &= 3726\,\mathrm{W}\cdot \left(\frac{T_\mathrm{4,top}-T_\mathrm{3,mid}}{\mathrm{K}}\right),
\end{align}
with $Q_{\mathrm{H}}$ being the heater power applied depending on the selected reflux in the distillation column.
Furthermore, the following relation is used:
\begin{equation}
    Q_{\mathrm{Cu,bot}} = Q_{\mathrm{f}} + Q_{\mathrm{no-fins}}.
\end{equation}
Applying equations\,\ref{eq:film2}-\ref{eq:Qfins} to the reboiler conditions and using xenon properties at a GXe temperature of $T_5 = 190$\,K taken from table \ref{tab:gas_values} yields
\begin{align}  \label{eq:film2,reb}                              
    h_\mathrm{Cu-GXe}(\Delta T_\mathrm{Cu-GXe} ) &= 2660  \cdot \left( \frac{\mathrm{K}}{T_5-T_4} \right)^{1/4} \,\frac{\mathrm{W}}{\mathrm{m}^2\,\mathrm{K}} \\
    Q_\mathrm{no-fins} &= 242  \cdot  \left(\frac{T_5-T_\mathrm{4\,top}}{\mathrm{K}} \right)^{3/4}\,\mathrm{W} \\
    \frac{d^2T_4}{dz^2}(z) &= -2188  \left(\frac{T_5-T_4(z)}{\mathrm{K}} \right)^{3/4} \frac{\mathrm{K}}{\mathrm{m^2}} \\
    T_4(z) &= - 2.33 \cdot 10^6 \cdot (z-z_0)^8 \,\frac{\mathrm{K}}{\mathrm{m^8}} +T_5 \\
    Q_\mathrm{f}(z) &=  5.19 \cdot 10^8 \cdot \left( z-z_0\right)^{7} \frac{\mathrm{W}}{\mathrm{m^7}}.
\end{align}   
Figure \ref{fig:reboiler_design_profile} shows the maximum heat transfer across the reboiler for the design conditions. A LXe temperature of 178\,K in the top reboiler part is assumed corresponding to the column's design operation pressure of 2\,bar. In the bottom reboiler vessel, a GXe temperature of 190\,K is assumed, corresponding to the design pressure of 3.5\,bar provided by the compressor. Furthermore, a heating power of $Q_\mathrm{H} = 925$\,W is assumed. The matching heat transfer at $T_\mathrm{4,top} = T_\mathrm{4,top2}$ was calculated to be $Q_{\mathrm{Cu,top}} = 3293$\,W and thus, $Q_{\mathrm{Cu,bot}} = 2368$\,W. This is a factor 1.3 larger than the required heat transfer for liquefaction in the bottom. The temperature difference $\Delta T_\mathrm{LXe-Cu}$ is computed to be 9\,K falling into the range for ONB measured in Ref.\,\cite{yy}.

\begin{figure}[h!]
\centering 
\includegraphics[width=0.8\textwidth]{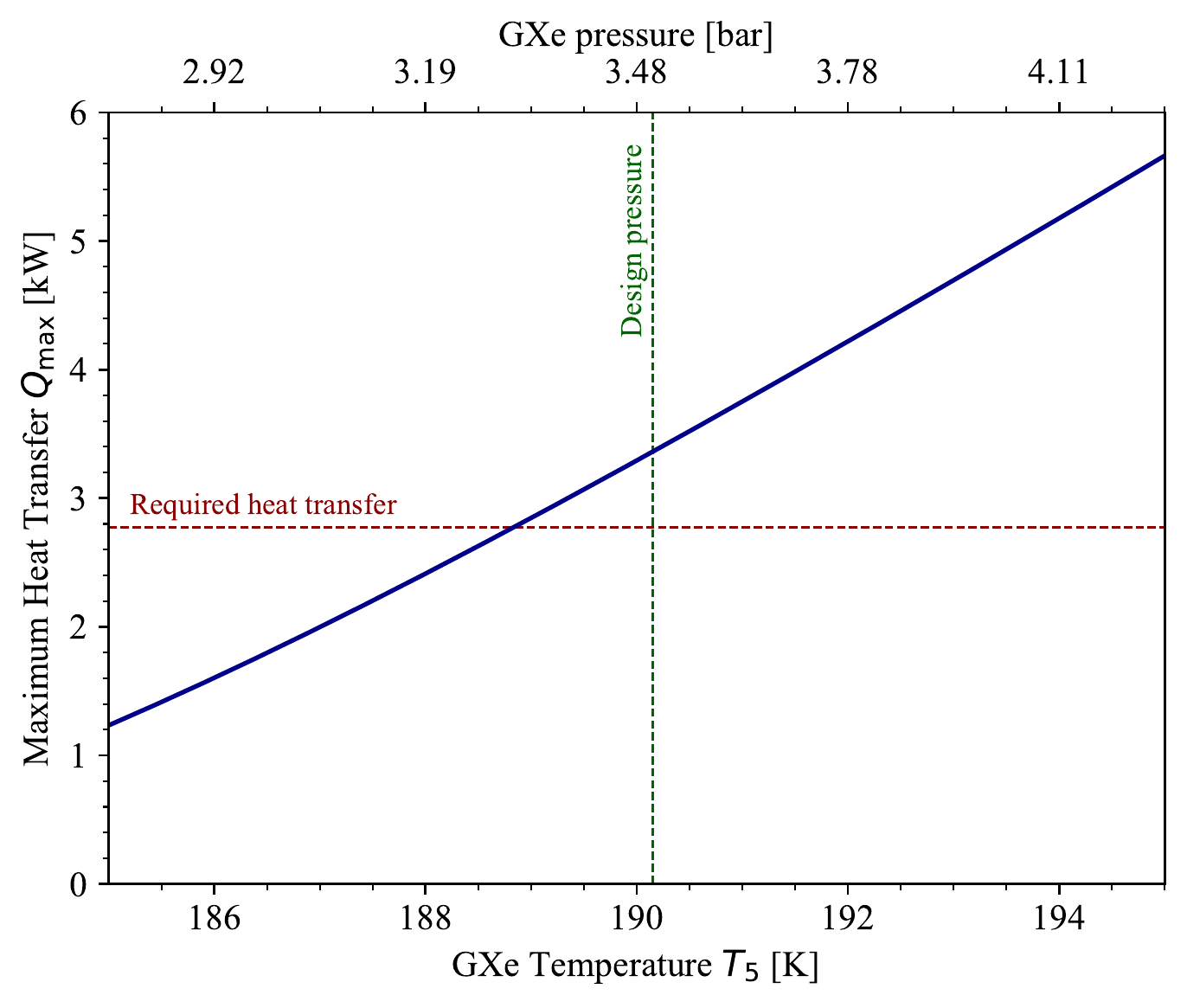}
\caption{Heat transfer $Q_\mathrm{max}$ as a function of $T_\mathrm{5}$ (bottom) and GXe pressure (top) in the bottom reboiler part. The horizontal dashed line shows the required heat transfer at the top reboiler part, while the vertical dashed line indicates the design pressure provided by the compressor.}
\label{fig:Q_function_T5_reb} 
\end{figure}

We can further investigate the expected heat transfer as a function of $T_5$ for the case of different performances of the compressor. For that, we repeated the calculations above for a range of 185\,K $\leq T_5 \leq$ 195\,K corresponding to a pressure range of about 2.8\,bar to 4.3\,bar deriving the xenon properties from Ref.\,\cite{rr}. The results are presented in figure\,\ref{fig:Q_function_T5_reb}.
The measurements in Ref.\,\cite{u} show that the compressor used can provide GXe pressures well above 4\,bar allowing to enhance the expected heat transfer by a factor of two and above compared to the required heat transfer.
As the liquefaction capabilities of the reboiler can only be tested when running the fully commissioned radon distillation column, the result of its performance are presented in a dedicated publication along with more details about the thermodynamic concept \cite{s}.
\newpage
\section{Conclusion}\label{VI}
Customized cryogenic heat exchangers are needed for ultra-pure noble gas applications, in this case for the construction of a high-flow radon distillation system for the direct dark matter search experiment XENONnT. Therefore, we designed bath-type heat exchangers featuring high conductivity copper in the form of large-area fins. 

The prototype HE with a diameter of 30\,cm acts as the radon column's top condenser and is liquefying xenon with liquid nitrogen. For the column's reflux ratio of 0.5, a xenon liquefaction rate of 36\,kg/h is needed corresponding to a cooling power of 925\,W for a xenon saturation temperature of 178\,K. A conservative model was developed to compute the expected heat transfer for the given design conditions assuming film boiling in the liquid nitrogen part and film condensation in the xenon part. The calculations yield an expected heat transfer of 1380\,W, well above the requirements. Measurements with this heat exchanger, while already installed as the top condenser, demonstrated a cooling power larger 3\,kW limited only by the measurement procedure. A cooling efficiency of $(0.98\pm 0.03)$ was derived from these tests. The achieved xenon liquefaction rate of more than 113\,kg/h is more than a factor of 3 above the design requirements allowing for a potential up-scaling of the radon distillation column's process flow.

Based on the experience with the prototype, the main challenge was to design a 50\,cm diameter heat exchanger to liquefy xenon with xenon at the radon distillation column's reboiler. To achieve that, the purified gaseous xenon from the radon column is pressurized by a compressor and is then thermally connected with the liquid xenon storage of the radon distillation column. The relatively small temperature difference between the two xenon phases with respect to liquid nitrogen and xenon is the main challenge. In our design, a copper plate equipped with fins was electron-welded into the flange connecting the two xenon reservoirs. This guarantees absolute leak tightness between both vessels and contains the purity of radon-depleted xenon. For the reboiler heat exchanger, a xenon liquefaction rate of 72\,kg/h in the bottom reboiler is needed corresponding to a cooling power of 1850\,W, while a xenon evaporation rate of 108\,kg/h in the top reboiler is required corresponding to a heating power of 2775\,W. Therefore, additional heat is introduced via electrical heaters in the copper plate. The developed model was applied to derive the expected heat transfer for the given reboiler design conditions, where we assume pool nucleate boiling between the top copper plate surface and the liquid xenon as well as film condensation between the bottom copper surface and the gaseous xenon. For a xenon saturation pressure of 2\,bar (178\,K) in the top and 3.5\,bar (190\,K) in the bottom, the heat transfer is computed to be 3293\,W fulfilling the requirements. The compressor we built can even provide pressures above 4\,bar yielding expected heat transfers larger than a factor of 2 above the design. The reboiler perfomance is published in a dedicated radon distillation column commissioning paper.
\newpage\clearpage

\acknowledgments
The heat exchanger development at Muenster University has been supported by the German Ministery for Education and Research (BMBF) under numbers
05A17PM2 and 05A20PM1. We acknowledge the contributions by the Electrical and the Precision Mechanical Workshop of the Institute for Nuclear Physics at University of M\"unster. We thank Henning Schulze-Eißing and Axel Buss for the pictures taken of our apparatus.



\begin{thebibliography}{99}

\bibitem{aa}
E. Aprile and T. Doke,
\emph{Liquid Xenon Detectors for Particle Physics and Astrophysics},\\
Rev. Mod. Phys. 82: 2053-2097 (2010).

\bibitem{aaa}
ICARUS Collaboration,
\emph{Operation and performance of the ICARUS-T600 cryogenic plant at Gran Sasso underground Laboratory},
JINST 10 P12004 (2015).

\bibitem{aaaaaa}
MicroBoNE Collaboration,
\emph{Measurement of the Flux-Averaged Inclusive Charged-Current Electron Neutrino and Antineutrino Cross Section on Argon using the NuMI Beam and the MicroBooNE Detector},
arXiv:2101.04228 (2021).

\bibitem{aaaa}
DUNE Collaboration,
\emph{First results on ProtoDUNE-SP liquid argon time projection chamber performance from a beam test at the CERN Neutrino Platform},
JINST 15 P12004 (2020).

\bibitem{aaaaa}
DUNE Collaboration,
\emph{Deep Underground Neutrino Experiment (DUNE) Near Detector Conceptual Design Report},
arXiv:2103.13910 (2021).

\bibitem{a}
DarkSide Collaboration,
\emph{DarkSide-20k: a 20 Tonne two-phase LAr TPC for direct dark matter detection at LNGS},
Eur. Phys. J. Plus 133: 131 (2018).

\bibitem{f}
XENON Collaboration, 
\emph{The XENON1T Dark Matter Experiment},
Eur. Phys. J. C 77: 881 (2017).

\bibitem{g}
XENON Collaboration, 
\emph{Projected WIMP Sensitivity of the XENONnT Dark Matter Experiment},\\
JCAP 11, 031 (2020).

\bibitem{h}
PandaX Collaboration,
\emph{PandaX: A Liquid Xenon Dark Matter Experiment at CJPL},\\
Science China Physics, Mechanics \& Astronomy, 57, 1476–1494 (2014).

\bibitem{hh}
PandaX Collaboration,
\emph{Dark Matter Search Results from the PandaX-4T Commissioning Run},\\
arXiv:2107.13438 (2021).

\bibitem{i}
LZ Collaboration,
\emph{The LUX-ZEPLIN (LZ) Experiment},
NIMA A 953 0168-9002 (2019).

\bibitem{ii}
DARWIN Collaboration
\emph{DARWIN: towards the ultimate dark matter detector},\\
JCAP 1611 no.11, 017 (2016).

\bibitem{b}
V.D. Ashitkov, A.S. Barabash et al.,
\emph{Liquid Argon Ionization Detector for Double Beta Decay Studies},
arXiv:nucl-ex/0309001 (2003).

\bibitem{c}
EXO Collaboration, 
\emph{Search for Neutrinoless Double-Beta Decay with the Upgraded EXO-200 Detector},
Phys. Rev. Lett. 120, 072701 (2018).

\bibitem{d}
NEXO Collaboration, 
\emph{Sensitivity and discovery potential of nEXO to neutrinoless double beta decay},
Phys. Rev. C 97, 065503 (2018).

\bibitem{dd}
NEXO Collaboration, 
\emph{nEXO: Neutrinoless double beta decay search beyond 1028 year half-life sensitivity}
arXiv:2106.16243 (2021).

\bibitem{e}
NEXT Collaboration, 
\emph{Sensitivity of a tonne-scale NEXT detector for neutrinoless double beta decay searches}, FERMILAB-PUB-20-299-ND-SCD, arXiv:2005.06467 (2020).


\bibitem{j}
MEG Collaboration,
\emph{The MEG detector for $\mu+$ → e + $\gamma$ decay search},\\
Eur. Phys. J. C, 73: 2365 (2013).

\bibitem{jj}
C. Romo-Luque on behalf of the PETALO Collaboration,
\emph{PETALO: Time-of-Flight PET with liquid xenon},
Nucl. Instr. Meth. A 958 162397 (2020).

\bibitem{jjj}
Y. Xing et al,
\emph{XEMIS: Liquid Xenon Compton Camera for 3$\gamma$ Imaging},\\
 Springer Proc.Phys./213 154 (2018).

\bibitem{n}
XENON Collaboration, 
\emph{Material radiopurity control in the XENONnT experiment},\\
arXiv:2112.05629 (2021).

\bibitem{k}
XENON Collaboration,
\emph{$^{222}$Rn emanation measurements for the XENON1T experiment},\\
Eur. Phys. J. C, 81: 337 (2021).

\bibitem{l}
EXO Collaboration,
\emph{Investigation of radioactivity-induced backgrounds in EXO-200},\\
Phys. Rev. C, 92 015503 (2015).

\bibitem{m}
NEXT Collaboration,
\emph{Radon and material radiopurity assessment for the NEXT double beta decay experiment},
AIP Conference Proceedings 1672, 060002 (2015).

\bibitem{o}
XMASS Collaboration, 
\emph{Radon removal from gaseous xenon with activated charcoal},\\
NIMA A 661 50 57 (2011).

\bibitem{p}
E. H. Miller et al., 
\emph{Constraining radon backgrounds in LZ},\\
AIP Conference Proceedings, 1921, 050003 (2018).

\bibitem{q}
XENON Collaboration, 
\emph{Online Rn-222 removal by cryogenic distillation in the XENON100 experiment},
Eur. Phys. J. C, 77: 358 (2017).

\bibitem{s}
M. Murra, D. Schulte, et al., 
\emph{Design and construction of a high-flow radon removal system for XENONnT},
under preparation.

\bibitem{r}
XENON Collaboration, 
\emph{Radon removal system of XENONnT},
under preparation.

\bibitem{rr}
National Institute of Standards and Technology. (Department of Commerce, Washington, D.C.), 
\emph{Data from NIST Standard Reference Database 69: NIST Chemistry WebBook}.

\bibitem{rs}
 	E. Baly, F. Donnan, \emph{The variation with temperature of the surface energies and densities of liquid oxygen, nitrogen, argon, and carbon monoxide}, 
 	J.Chem.Soc.London 81 (1902) 907-923.
 	
\bibitem{rrr}
Bumax AB,
\emph{Bumax 88 datasheet},  [online],\\
https://www.bumax-fasteners.com/de/technische-informationen/datenblatter (Accessed Oct. 14, 2021)

\bibitem{rrrr}
P. Bradley, R. Radebaugh, (2013), Properties of Selected Materials at Cryogenic Temperatures, CRC Press, Boca Raton, FL, [online], https://tsapps.nist.gov/publication/get\_pdf.cfm?pub\_id=913059 (Accessed Feb. 1, 2022).





\bibitem{ss}
F. Incropera, D. De Witt, et al.,
\emph{Fundamentals of Heat and Mass Transfer},
Wiley, 6th. ed. (2007).


\bibitem{u}
D. Schulte, et al., 
\emph{Ultra-clean radon-free four cylinder magnetically-coupled piston pump},\\
2021 JINST 16 P09011.

\bibitem{v}
R. Barron, 
\emph{Cryogenic heat transfer},
Taylor \& Francis, 1999.

\bibitem{vv}
L. Bromley, 
\emph{Heat transfer in stable film boiling},
California Digital Library, 1949.

\bibitem{w}
P. Berenson, 
\emph{Transition boiling heat transfer from a horizontal surface},\\
MIT Libraries, Technical Report No. 17, 1960.

\bibitem{x}
N. Zuber, 
\emph{Further remarks on the stability of boiling heat transfer},\\
United States Atomic Energy Commission, Report 58.-5. Project 34, 1958.

\bibitem{y}
T. Haruyama,
\emph{Boiling heat transfer characteristics of liquid xenon},\\
AIP Conference Proceedings 613, 1499 (2002) 

\bibitem{yy}
P.A. Breur, et al.,
\emph{Measurement of the onset of nucleate boiling in liquid xenon},\\
2021 JINST 16 T02004.

\bibitem{yyy}
T. L. Bergman, A. S. Lavine,
\emph{Fundamentals of heat and mass transfer},
Wiley, 8th Edition 2018.



\end{thebibliography}
\end{document}